\documentclass[preprint]{aastex}
\usepackage{color,natbib,lscape}

\newcommand{\HST}{{\sl HST}}

\newcommand{\Lsun}{\mbox{$L_{\sun}$}}

\newcommand{\Mjup}{\mbox{$M_{\rm Jup}$}}

\newcommand{\degree}{\mbox{$^{\circ}$}}

\newcommand{\etal}{et al.}
\newcommand{\eg}{e.g.}
\newcommand{\ie}{i.e.}
\newcommand{\cf}{cf.}

\newcommand{\kms}{\hbox{km~s$^{-1}$}}

\newcommand{\htwoo}{{\hbox{H$_2$O}}}   % H20
\newcommand{\htwo}{{\hbox{H$_2$}}}     % H2
\newcommand{\meth}{{\hbox{CH$_4$}}}   % CH4
\newcommand{\ammonia}{{\hbox{NH$_3$}}}   % NH3

\newcommand{\Ks}{\mbox{$K_S$}}
\newcommand{\degs}{\mbox{$^{\circ}$}}
\newcommand{\Lbol}{\mbox{$L_{\rm bol}$}}

\newcommand{\Teff}{\mbox{$T_{\rm eff}$}}
\newcommand{\logg}{\mbox{$\log(g)$}}

\newcommand{\methshort}{\mbox{$CH_4s$}}

\newcommand{\cfbdsAB}{\hbox{CFBDSIR~J1458+1013AB}}
\newcommand{\cfbdsA}{\hbox{CFBDSIR~J1458+1013A}}
\newcommand{\cfbdsB}{\hbox{CFBDSIR~J1458+1013B}}
\newcommand{\cfbds}{\hbox{CFBDSIR~J1458+1013}}
\newcommand{\ugps}{\hbox{UGPS~J0722$-$05}}

\slugcomment{ApJ, in press (accepted March 4, 2011)}
\shorttitle{A Very Cold Substellar Binary}
\shortauthors{Liu \etal}

\begin{document}

\title{CFBDSIR~J1458+1013B:\\A Very Cold ($>$T10) Brown Dwarf in a Binary System\altaffilmark{1,2,3}}
%\title{CFBDSIR~J1458+1013AB:\\A Very Cold (T9.5 + $>$T10) Substellar
%  Binary\altaffilmark{1,2,3}}


\author{Michael C. Liu,\altaffilmark{4}, 
  Philippe Delorme,\altaffilmark{5} 
  Trent J. Dupuy,\altaffilmark{6,7} 
  Brendan P. Bowler,\altaffilmark{4} \\
  Loic Albert,\altaffilmark{8} 
  Etienne Artigau,\altaffilmark{9} 
  Celine Reyl\'e,\altaffilmark{10}
  Thierry Forveille,\altaffilmark{5} 
  Xavier Delfosse\altaffilmark{5}}

\altaffiltext{1}{Some of the data presented herein were obtained at the
  W.M. Keck Observatory, which is operated as a scientific partnership
  among the California Institute of Technology, the University of
  California, and the National Aeronautics and Space Administration. The
  Observatory was made possible by the generous financial support of the
  W.M. Keck Foundation.}
\altaffiltext{2}{Based on observations made with ESO Telescopes at the
  Paranal Observatory under programme 085.C-0805(A).}
\altaffiltext{3}{Based on observations obtained with WIRCam, a joint
  project of CFHT, Taiwan, Korea, Canada, France, and the
  Canada-France-Hawaii Telescope (CFHT) which is operated by the
  National Research Council (NRC) of Canada, the Institute National des
  Sciences de l'Univers of the Centre National de la Recherche
  Scientifique of France, and the University of Hawaii.}
%----------------------------------------%
\altaffiltext{4}{Institute for Astronomy, University of Hawai`i, 2680
  Woodlawn Drive, Honolulu, HI 96822, USA; mliu@ifa.hawaii.edu.}
\altaffiltext{5}{UJF-Grenoble 1 / CNRS-INSU, Institut de Plan\'etologie
  et d’Astrophysique de Grenoble (IPAG) UMR 5274, Grenoble F-38041,
  France.}
\altaffiltext{6}{Harvard-Smithsonian Center for Astrophysics, 60 Garden Street,
Cambridge, MA 02138}
\altaffiltext{7}{Hubble Fellow}
\altaffiltext{8}{Canada-France-Hawaii Telescope Corporation, 65-1238
  Mamalahoa Highway, Kamuela, HI 96743, USA}
\altaffiltext{9}{D\'epartement de Physique and Observatoire du Mont
  M\'egantic, Universit\'e de Montr\'eal, C.P. 6128, Succ. Centre-Ville,
  Montréal, QC H3C 3J7, Canada.}
\altaffiltext{10}{Observatoire de Besan\c{c}on, Universit\'e de
  Franche-Comt\'e, Institut Utinam, UMR CNRS 6213, BP 1615, 25010
  Besan\c{c}on Cedex, France.}

\begin{abstract} 
  \noindent We have identified \cfbds\ as a 0.11\arcsec\ (2.6~AU)
  physical binary using Keck laser guide star adaptive optics imaging
  and have measured a distance of $23.1\pm2.4$~pc to the system based on
  near-IR parallax data from CFHT.
  The integrated-light near-IR spectrum indicates a spectral type of
  T9.5, and model atmospheres suggest a slightly higher temperature and
  surface gravity than the T10~dwarf \ugps.
  Thus, \hbox{\cfbdsAB} is the coolest brown dwarf binary found to date.
  Its secondary component has an absolute $H$-band magnitude that is
  1.9$\pm$0.3~mag fainter than \ugps, giving an inferred spectral type
  of $>$T10. The secondary's bolometric luminosity of
  $\sim2\times10^{-7}$~\Lsun\ makes it the least luminous known brown
  dwarf by a factor of 4--5.
  By comparing to evolutionary models and T9--T10 objects, we estimate a
  temperature of 370$\pm$40~K and a mass of 6--15~\Mjup\ for \cfbds{B}.
  At such extremes, atmospheric models predict the onset of novel
  photospheric processes, namely the appearance of water clouds and the
  removal of strong alkali lines, but their impact on the emergent
  spectrum is highly uncertain.
  Our photometry shows that strong \meth\ absorption persists at
  $H$-band; the $J-K$ color is bluer than the latest known T~dwarfs but
  not as blue as predicted by current models; and the $J-H$ color
  delineates a possible inflection in the blueward trend for the latest
  T~dwarfs.
  Given its low luminosity, atypical colors and cold temperature,
  \cfbds{B} is a promising candidate for the hypothesized Y~spectral
  class. However, regardless of its ultimate classification, 
  \cfbdsAB\ provides a new benchmark for measuring the properties of
  brown dwarfs and gas-giant planets, testing substellar models, and
  constraining the low-mass limit for star formation.
\end{abstract}


%----------------------------------------------------------------------%

\section{Introduction}

Over the last 15 years, stellar and exoplanetary studies have been
propelled through the study of brown dwarfs. Residing at the extremes of
low mass, luminosity and temperature, brown dwarfs serve as laboratories
for understanding gas-giant extrasolar planets as well as the faint end
of the star formation process. The coolest known brown dwarfs, the
T~dwarfs, have temperatures ($\approx$600--1400~K), bolometric
luminosities ($\approx10^{-4}$ to $10^{-6}$~\Lsun), molecule-dominated
(\htwoo\ and \meth) IR spectra, and atmospheric processes (\eg,
non-equilibrium chemistry) that are more akin to Jupiter than any star.
This physical similarity drives the ongoing search for even more extreme
objects, to close the gap of $\approx$400~K in temperature and factor of
$\approx$1000 in \Lbol\ between the coolest T~dwarfs and the planet
Jupiter. 
Along the way, one question to be answered is whether such new
discoveries will have sufficiently distinct spectra to invoke the
proposed ``Y'' spectral type \citep{1999ApJ...519..802K}.

A well-established strategy for finding the coolest objects is to search
for them as companions to other objects. This approach has been quite
successful: the first L~and T~dwarfs were found as companions around
hotter objects, GD~165B \citep{1988Natur.336..656B} and Gl~229B
\citep{1995Natur.378..463N}, respectively. To date, some of the coolest
T~dwarfs are found as companions to stars: Gl~570~D (T7.5;
\citealp{2000ApJ...531L..57B}), HD~3651~B (T7.5;
\citealp{2006astro.ph..8484M, 2006astro.ph..9464L}), ULAS~J1416+13~B
(T7.5; \citealp{2010A&A...510L...8S, 2010MNRAS.404.1952B}), GJ~758B
($\approx$T8--T9; \citealp{2009ApJ...707L.123T, 2011ApJ...728...85J})
Wolf~940B (T8.5; \citealp{burningham08-T8.5-benchmark}), and Ross~458C
(T8.5; \citealp{2010MNRAS.405.1140G, 2010A&A...515A..92S}).

Field L~and T~dwarfs themselves have long been recognized as promising
search targets for even cooler objects. 
To date, six secondary components later than~T5 have been identified in
binaries (see compilation in \citealp{2010liu-2m1209}), one of which has
a short enough orbital period to yield a dynamical mass, the T5.0+T5.5
substellar binary 2MASS~J1534$-$2952AB
\citep{liu08-2m1534orbit}.\footnote{The T8~dwarf 2MASS~J0939$-$24 has
  also been suggested to be an unresolved binary based on its absolute
  spectrophotometry \citep{2008ApJ...689L..53B, 2009ApJ...695.1517L}.}
In addition, discovery of ultracool objects in binaries opens the door
to direct mass measurements through astrometric and spectroscopic
monitoring of their orbits.

As part of an ongoing program to study the multiplicity and physical
properties of brown dwarfs using laser guide star adaptive optics (LGS
AO), we present here the discovery of the binarity of \cfbds. This
object was identified in deep $z$ and $J$-band data by
\citet{2010A&A...518A..39D} and assigned a preliminary spectral type of
T8.5 based on its $H$-band spectrum. We have scrutinized this object
using a combination of high resolution imaging, near-IR astrometry, and
near-IR spectroscopy. These data indicate an integrated-light spectral
type of T9.5, which is later than all known field objects except the
T10~dwarf \ugps\ \citep{2010MNRAS.408L..56L}. The secondary component of
\cfbdsAB\ is $\approx$1.5--2~mag fainter in absolute magnitude and
$\approx$150~K cooler in temperature than any known brown dwarf,
making it a promising candidate for the first member of the Y~spectral
class.

%----------------------------------------------------------------------%
\section{Observations}

\subsection{Keck LGS AO Imaging and Photometry}

We imaged \cfbds\ on UT~22~May~2010 and 08~July~2010 using the sodium
LGS AO system of the 10-meter Keck II Telescope on Mauna Kea, Hawaii
\citep{2006PASP..118..297W, 2006PASP..118..310V}. We used the facility
IR camera NIRC2 with its narrow field-of-view camera, which produces a
$10.2\arcsec \times 10.2\arcsec$ field of view. Conditions were
photometric for both runs with excellent seeing: 0.5--0.6\arcsec\ and
0.3--0.5\arcsec\ FWHM in May and July, respectively, as measured by a
differential image motion monitor located at the nearby
Canada-France-Hawaii Telescope. The LGS provided the wavefront reference
source for AO correction, with the tip-tilt motion measured
contemporaneously from the $R=15.5$~mag field star
USNO-B1.0~1002--0236935 \citep{2003AJ....125..984M} located 55\arcsec\
away from \cfbds.
The LGS brightness, as measured by the flux incident on the AO wavefront
sensor, was equivalent to a $V\approx9.6$ and 10.3~mag star in May and
July, respectively.

At each epoch, we obtained a series of dithered images by offsetting the
telescope by a few arcseconds between each 1--2 images. The sodium laser
beam was pointed at the center of the NIRC2 field-of-view for all
observations.
In May~2010, we successfully resolved \cfbds\ into a tight binary using
the moderate-band $CH_4s$ filter, which has a central wavelength of
1.592~\micron\ and a width of 0.126~\micron.
In July~2010, we obtained images with the $J$ (1.25~\micron), $H$
(1.64~\micron), and $K$ (2.20~\micron) filters from the Mauna Kea
Observatories filter consortium \citep{mkofilters1, mkofilters2}.

The images were analyzed in a similar fashion as our previous LGS papers
\citep[e.g.][]{2006astro.ph..5037L, liu08-2m1534orbit}. The raw images
were reduced using standard methods of flat-fielding and
sky-subtraction. After rejecting individual images of poor quality, we
registered and stacked the data into a final mosaic. The binary's flux
ratios and relative astrometry were derived by fitting an analytic model
of the point spread function (PSF) as the sum of two elliptical
gaussians to the images of the binary in the mosaics.
We did not correct the relative astrometry for instrumental optical
distortion, as the size of the effect as predicted from a distortion
solution by B. Cameron (priv. comm.) is insignificant compared to our
measurements uncertainties.
We adopted a pixel scale of $9.963 \pm 0.005$~mas/pixel and an
orientation for the detector's $+y$~axis of $+0.13 \pm 0.07$\degs\ for
NIRC2 \citep{2008ApJ...689.1044G}.

\begin{figure}
\vskip -4in
\centerline{\includegraphics[height=6.5in,angle=90]{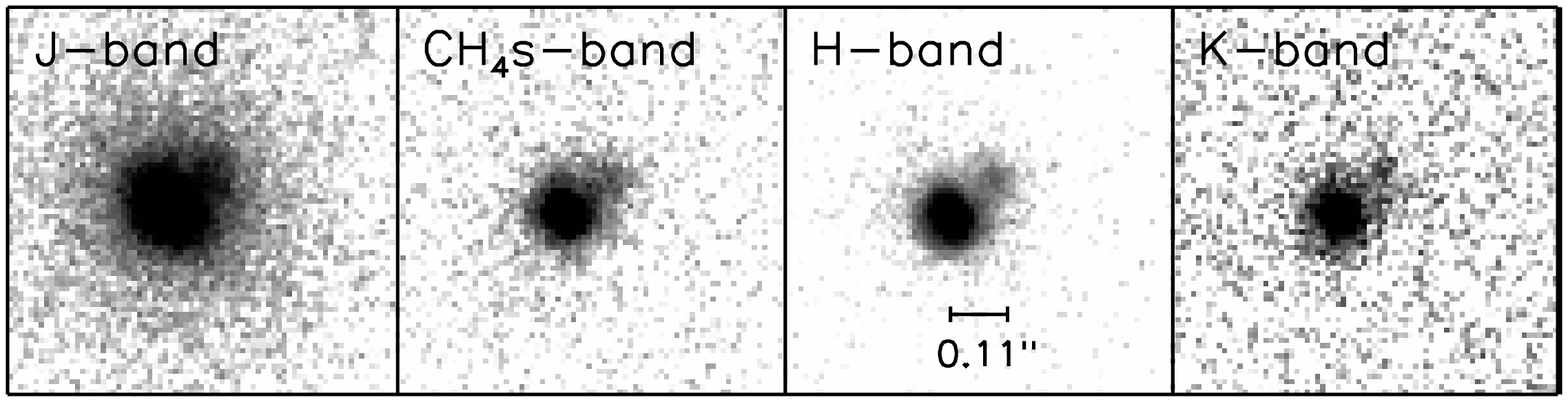}}
%\centerline{\includegraphics[height=6.5in,angle=90]{make-figure-cfbds1458.ps}}
\vskip -3in
\centerline{\includegraphics[height=6.5in,angle=90]{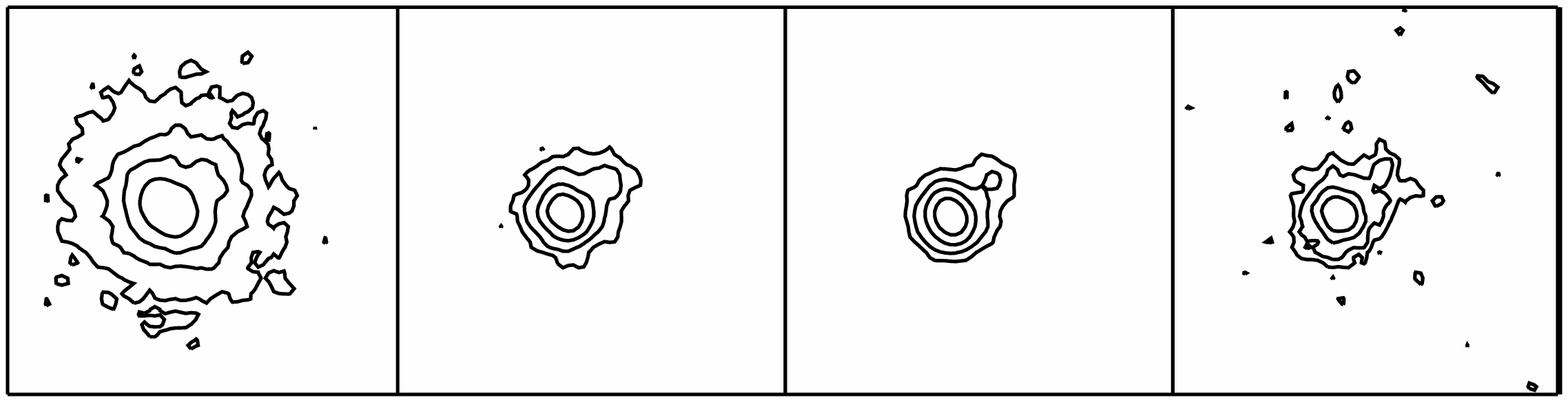}}
%\centerline{\includegraphics[height=6.5in,angle=90]{make-figure-contours-cfbds1458.ps}}
%\vskip -2ex
\caption{\normalsize Keck LGS AO imaging of \cfbdsAB. North is up and
  east is left. Each image is 0.75\arcsec\ on a side. The greyscale
  images are shown with a power-law stretch with an exponent of 1/4. The
  contours are shown after smoothing the images by a 2-pixel gaussian.
  Contours are drawn from 90\%, 45\%, 22.5\%, 11.2\%, and 5.6\% of the
  peak value in each bandpass. \label{fig:images}}
\end{figure}

To validate our measurements, we created myriad artificial binary stars
from the science data themselves; given the large flux ratio of the
binary, a scaled and shifted version of the primary star provides an
excellent simulation of the data \citep[e.g.][]{2009ApJ...699..168D}.
For each filter, our fitting code was applied to artificial binaries
with similar separations and flux ratios as \cfbdsAB\ over a range of
position angles (PAs). The RMS scatter of the simulations were adopted
as our measurement errors; the results also showed that any systematic
offsets are small. The exceptions to this were the $J$ and $H$-band
separation measurements, where the average systematic offset from the
simulations was slightly larger than the RMS scatter, and for these
datasets we adopted the offset as the uncertainty.
Table~\ref{table:keck} presents our final Keck LGS measurements, and
Figure~\ref{fig:images} shows our Keck LGS images.

We use our measured flux ratios and the published $JH\Ks$ photometry
from \citet{2010A&A...518A..39D} to derive resolved IR colors and
magnitudes for \cfbdsAB\ on the MKO system.
Delorme \etal\ report data using the \Ks-band filter while our Keck LGS
imaging used the $K$-band filter; to compute resolved $K$-band
photometry, we synthesize an integrated-light color of $(K-\Ks) =
-0.17\pm0.02$~mag from the published spectra of ULAS~J1238+09 (T8.5),
Wolf~940B (T8.5), ULAS~J0034$-$00 (T9), CFBDS~J0059$-$01 (T9), and
ULAS~J1335+11 (T9) \citep{2008MNRAS.391..320B,
  burningham08-T8.5-benchmark, 2007MNRAS.381.1400W,
  2008A&A...482..961D}. This is consistent with the color synthesized
from the \cfbdsAB\ spectrum itself (\S~\ref{sec:spectrum}), but the
signal-to-noise (S/N) in the $K$-band spectrum is quite low, so we use
the average from the known T8.5 and T9 objects.
To obtain the resolved $CH_4s$ photometry from the published $H$-band
data, we synthesizes a color \hbox{$(CH_4s-H) = -0.47\pm0.03$~mag}
directly from the integrated-light spectrum of \cfbdsAB.
Table~\ref{table:binary} presents the resolved photometry.

% - - - - - - - - - - - - - - - - - - - - - - - - %

\subsection{CFHT Infrared Astrometry}

To determine the distance to \cfbdsAB, we acquired seven epochs of
$J$-band imaging in 2009 and 2010 spanning 1.07 years using the
Canada-France-Hawaii 3.6-meter Telescope on Mauna Kea, Hawai`i. We used
the facility wide-field near-IR camera WIRCam
\citep{2004SPIE.5492..978P}, which has four 2048$\times$2048 Hawaii-2RG
detectors, each with an approximate field-of-view of $10\arcmin \times
10\arcmin$. At each epoch, we placed the target near the center of the
northeast detector.
The data were taken with slightly different observing procedures as they
originated from two independent programs: at 3 of the epochs, images of
60-s $\times$ 6 coadds were obtained at 10 dither positions, and at the
other 4 epochs, single 60-s images were obtained at 11 dither positions.
The median FWHM of the data was 0$\farcs$73 (min/max of
0$\farcs$55/0$\farcs$80). The target was always observed near transit,
with a maximum deviation in airmass of 0.03. This reduces any systematic
offsets in the target's astrometry from the differential chromatic
refraction between the late-T dwarf science target and the background
stars to $<1$~mas (Dupuy \& Liu, in preparation), well below the
measurement noise.

Our WIRCam astrometry pipeline is described in detail by
\citet{dupuy2010-PhD} and Dupuy \& Liu (in prep) and successfully
generates parallaxes that are in accord with previously published
results for L~and T~dwarfs.
We use SExtractor \citep{1996A&AS..117..393B} to generate an astrometric
catalog of all the sources in every individual image. We then
cross-match detections across dithered images and use the mean and rms
of each source to determine its final position and error at that epoch;
we exclude any source that has fewer than 7 detections or a S/N less
than~7.
We then cross-match these positions across different epochs, allowing
for a full linear transformation in the astrometry to account for scale
and orientation changes in the instrument. (WIRCam is installed and
removed from the telescope for each observing run.) After an initial
iteration, we remove any stars with large proper motions ($>50$~mas/yr)
or parallaxes ($>3\sigma$) from the epoch-matching fit. Finally, we
absolutely calibrate our astrometry by matching to 2MASS to determine
the overall scale, orientation, and shear. For this target, 8~of our
72~reference stars were well-detected in 2MASS and had sufficiently low
proper motion ($<30$~mas/yr) to be used for absolute calibration.

We then fit for the proper motion and parallax of every source in the
field. We measured a proper motion of $432\pm6$~mas/yr at a PA of
$154.2\pm0.7\deg$ for \cfbds, in good agreement with the
\citet{2010A&A...518A..39D} measurement of $444\pm16$~mas/yr and
$157.5\pm2.1\degs$ from seeing-limited images.
\hbox{\cfbds} was the only object we measured in the field with a
significant parallax, $42.4\pm4.5$~mas relative to the reference stars.
(To our knowledge, this is the faintest object outside the solar system
with an optical or IR parallax measurement, \cf,
\citealp{2010A&A...524A..38M}.) To account for the mean parallax of our
reference stars, we used the Besan\c{c}on model of the Galaxy
\citep{2003A&A...409..523R} to simulate the distance distribution of
sources in the direction of \cfbds\ within the magnitude range of our
images ($15.1 < J < 20.1$~mag). We find a mean parallax of
$0.9\pm0.1$~mas from the Galaxy model, with the error accounting for the
sampling variance due to the finite number of reference stars. This
results in an absolute parallax of $43.3\pm4.5$~mas for \cfbds\
($23.1\pm2.4$~pc), in good agreement with the original 23~pc photometric
estimate by \citet{2010A&A...518A..39D}. The resulting tangential
velocity (47$\pm$5~\kms) is higher than the median for field T~dwarfs,
but not extraordinary given the observed velocity dispersion of the
field objects \citep[e.g.][]{2004AJ....127.2948V, 2009AJ....137....1F}.
Table~\ref{table:astrometry} contains our astrometry results.

% - - - - - - - - - - - - - - - - - - - - - - - - %

\subsection{VLT/X-Shooter Spectroscopy \label{sec:spectrum}}

We obtained integrated-light near-IR spectra of \cfbdsAB\ with the
X-Shooter spectrograph on VLT-UT2 (Program 085.C-0805). The observations
were carried out in four 1-hour observing blocks (hereafter OB) from
May~5 to July~9, 2010, achieving a total exposure time on target of
11700~sec, split into 8 A-B nods on the slit of 2 x 732~sec each. The
slit width was 1.2\arcsec, and the seeing varied between
0.8--1.0\arcsec. We chose a large slit size, because direct acquisition
of the target was impossible and we had to acquire using a blind offset
from a reference source.

The spectra for each OB were reduced using the standard ESO X-Shooter
pipeline \citep{2010SPIE.7737E..56M}, which produced a 2-dimensional,
curvature-corrected spectrum of the NIR arm of X-Shooter
(1.0--2.5~\micron). No signal was detected in the UVB and visible arm of
the instrument. The trace was extracted using our own IDL procedures,
using gaussian boxes in the spatial dimension at each wavelength. A
similar gaussian extraction box was used at about 5 times the FWHM of
the trace to obtain the noise spectrum of the sky. The resulting
1-dimensional spectrum from each science target OB was then divided by
the spectra of telluric stars observed just before or after, reduced and
extracted using the same pipelines.

The resulting 4 spectra from all the OBs were then median combined to
produce the final science spectrum. Since the resulting spectrum (with
$R \equiv \lambda/\Delta\lambda \approx4300$) has low S/N and most
analyses of late-T dwarfs have used much lower resolution spectra, we
smoothed the spectra using a weighted average over 50 pixels in the
wavelength dimension (corresponding to 3~and 5~nm at $J$-band and
$K$-band, respectively), weighting by the inverse variance of the
extracted sky spectrum. This approach uses the full spectral resolution
of X-Shooter to give a much lower weight to wavelengths affected by
telluric emission lines, thus improving the S/N with respect to simple
unweighted binning using the average or median.

Given the large wavelength range covered by X-Shooter, seeing variation
in the spectral direction can led to wavelength-dependent slit losses.
Therefore, we flux-calibrated the spectrum using $JH\Ks$ photometry of
\cfbds\ from \citet{2010A&A...518A..39D} and $Y$-band data from the
UKIDSS Data Release~8 ($Y = 20.58 \pm 0.21$~mag, on the Vega system).
Following \citet{2008A&A...484..469D}, we applied small scaling
factors to each of the broadband wavelength ranges so that the
magnitudes derived from the \cfbds\ spectrum matched the photometry. The
final spectrum is shown in Figure~\ref{fig:spectra}.

%----------------------------------------------------------------------%

\section{Results}

\subsection{Proof of Companionship}

\begin{figure}
%\vskip -4in
\centerline{\includegraphics[height=6.5in,angle=90]{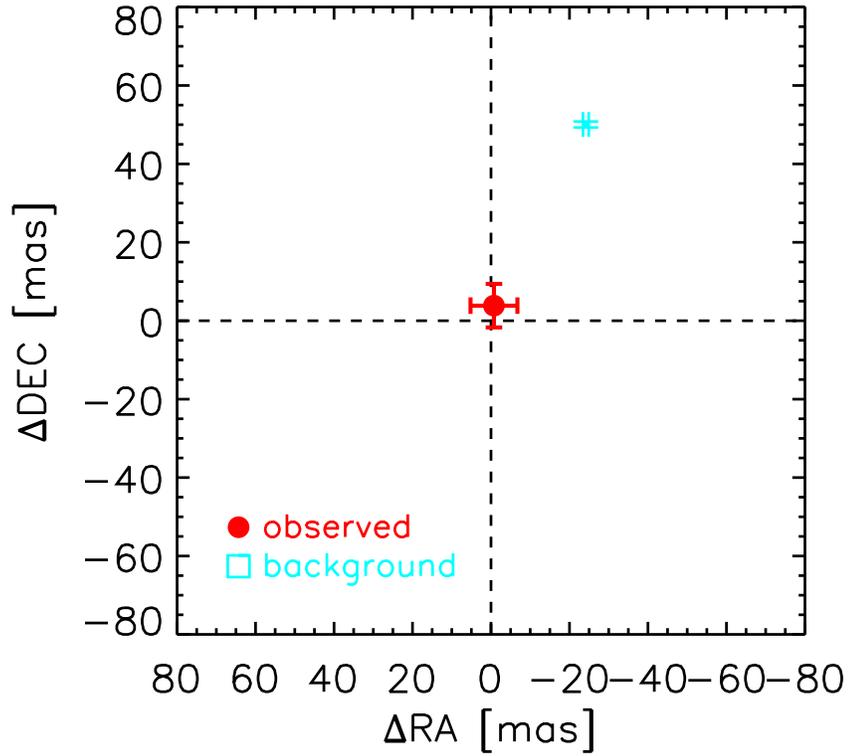}}
%\centerline{\includegraphics[height=6.5in,angle=90]{plot-background-object.ps}}
%\vskip -2ex
\caption{\normalsize The change in the position of \cfbds{B} relative to
  \cfbds{A}. The red point shows the measured change from our two Keck
  LGS epochs, which is consistent with no motion. The blue symbol shows the
  expected motion (and its uncertainties) if component~B is a background source, given the  
  measured parallax and proper motion of \cfbds. (Over the 1.5 month
  time baseline between Keck epochs, the proper motion dominates.) The
  two points are inconsistent at 9.1$\sigma$, indicating that the system
  is a physical binary.
  \label{fig:background}}
\end{figure}

Examination of the Digitized Sky Survey images shows no plausible
optical counterpart if component~B were a background object. Similarly,
the $z$-band imaging from July 2004 presented in
\citet{2010A&A...518A..39D} shows no star to a depth of $z\sim$24~mag,
2--3~mag deeper than expected for a background M~star if it had the
near-IR magnitude of component~B.
In fact, the high proper motion of the source means that our two Keck
epochs separated by 1.5 months are sufficient to confirm the two
components are physically associated. (For the July 2010 epoch, we
choose the Keck $H$-band astrometry, which has the smallest errors.)
Figure~\ref{fig:background} shows the large expected motion for
component~B if it were an object at infinite distance, given our
measured proper motion and parallax for \cfbds. (The proper motion
dominates over the parallax.) In contrast, the relative positions of the
two components do not change between our observations. The length of the
vector between the observed and predicted change is non-zero at
9.1$\sigma$. Thus, we conclude the two components are physically
associated.

%- - - - - - - - - - - - - - - - - - - - - - - - - - - - - -%

\subsection{Integrated-Light Spectrum  \label{sec:spectra}}

\subsubsection{Comparison with known late-T dwarfs}

The integrated-light spectrum of \cfbds{AB} is very similar to other
recent very late-T dwarf discoveries (Figure~\ref{fig:spectra}). 
As seen by its narrower continuum peak, its $J$-band spectrum shows
slightly stronger absorption than the T9~dwarfs CFBDS~J0059$-$01,
ULAS~J0034$-$00, and ULAS~J1335+11 and is very close to the T10~dwarf
\ugps.\footnote{\citet{2010MNRAS.408L..56L} report the presence of an
  unidentified, narrow absorption feature at 1.275~\micron\ in the
  spectrum of \ugps. They suggest it arises from \htwoo\ or HF.
  Figure~\ref{fig:spectra} suggests it may also be present in
  \cfbds{AB}. However, while the detection is significant in terms of
  S/N, it appears much broader than in the \ugps\ spectrum. Also, Lucas
  \etal\ reports a second absorption feature at 1.282~\micron, which
  coincides with a local peak in the \cfbds{AB} spectrum. Thus our
  spectra do not confirm either of these proposed features.} (We caution
that spectral types later than T8 are an active topic of investigation,
without a classification system established yet. For this discussion, we
simply adopt the published types for these objects, which may ultimately
change once a system is defined.)
In the $H$ band however, \cfbds\ is more similar to the warmer objects,
especially on the blue side of the peak. The absorption on the red side
of the $H$-band is the strongest observed in any known brown dwarf (see
the CH$_4$-H index in Table~\ref{table:indices}). This might be due to
enhanced collision induced absorption (CIA) of \htwo\ starting to
suppress the $H$-band peak; this notion fits well with the strongly
depressed $K$-band flux of \cfbds\ compared to the other late-T dwarfs,
where the strongest continuum absorber is CIA~\htwo.

The suppression could be due to either strong gravity or low
metallicity, and distinguishing between the two can be challenging
\citep[e.g.][]{2006liu-hd3651b}. Using model atmospheres,
\citet{2006ApJ...639.1095B} and \citet{2007ApJ...667..537L} show that
the shape of the $Y$-band continuum is more metallicity-dependent than
gravity-dependent (though their two sets of models show different
amplitudes for these effects). The main absorbers on the blue side of
the $Y$-band peak are the pressure broadened wings of the \ion{K}{1} and
\ion{Na}{1} fundamental doublets, which naturally decrease with lower
metallicity; more metal-rich models also have the $Y$-band peak at
redder wavelengths and the overall $Y$-band flux being fainter. The
observed $Y$-band spectra of \cfbds{AB} does not show any strong
differences with \ugps\ and CFBDS~J0059$-$01 (Figure~\ref{fig:spectra}),
though the peak might be slightly redder than the other two objects,
perhaps hinting at a greater metallicity (though this would be
discrepant with the highly suppressed $K$-band). Overall, we favor the
higher gravity explanation, though a better spectrum at shorter
wavelengths would help to investigate this further.

Taken altogether, the spectrum of \cfbds{AB} shows that this object is
slightly earlier than \ugps\ and possibly higher gravity, but later than
the known T9~dwarfs. Thus, we provisionally assign a spectral type of
T9.5, though of course it may turn out to be earlier or later type,
depending on the future development of classification for such cool
objects.

\subsubsection{Spectral index analysis with solar-metallicity models \label{sec:indices}}

To quantify the absorption features of \meth, \htwoo, CIA~\htwo, and
possibly NH$_3$, we calculated spectral indices previously used for
T~dwarfs \citep{2006ApJ...637.1067B, 2007MNRAS.381.1400W,
  2008A&A...482..961D}.
The measurements (Table~\ref{table:indices}) indicate that \cfbds{A} has
some of the strongest absorption features seen in the near-IR spectra of
late T dwarfs, exceeded by only \ugps\ for the $W_J$ and NH$_3$-H
indices. Given the $\approx$2~magnitudes flux difference between
component~A and~B, it is reasonable to consider the integrated spectrum
is dominated by the primary (as discussed below).

We use the solar-abundance BT-Settl-2009 models
\citep{2010HiA....15..756A} to estimate \Teff\ and \logg\ for
component~A from the index measurements. We compensate for the
systematic errors of the models \citep[e.g.][]{2007ApJ...667..537L,
  liu08-2m1534orbit} by scaling their index predictions to match those
measured for the T8.5 benchmark Wolf~940B
\citep{burningham08-T8.5-benchmark}.
Its high quality spectrophotometry, parallax, and age estimate from its
primary star enable a precise estimate of its properties using
evolutionary models (\Teff~=~575~K; \logg~=~4.75, [M/H]=0 from
\citealp{burningham08-T8.5-benchmark}; see also
\citealp{2010ApJ...720..252L}). With this approach, the temperatures and
gravities for the $>$T8 dwarfs derived from an index-index plot broadly
agree with the values derived from the analysis of their full near-IR
spectra \citep{2006ApJ...639.1095B, 2007MNRAS.381.1400W,
  2008A&A...482..961D, 2008MNRAS.391..320B, 2010MNRAS.408L..56L}.

Since the systematic errors in the models in this low-temperature regime
are likely much larger than the measurement errors, we use the model
grid only for relative estimates of temperature and gravity differences
among the late-T dwarfs, not absolute measurements.
Figure~\ref{fig:indices}
indicates that \cfbds{A} is as cool as \ugps\ and $\gtrsim$50~K or more
cooler than the other $>$T8 dwarfs. (The caveat is that the model
analysis assumes a single metallicity for all the objects, consistent
with the similarity in $Y$-band shapes discussed above.) The plot also
puts \cfbds{A} at a higher ($\gtrsim$0.5~dex) surface gravity than the
other objects, as expected from the highly suppressed $K$-band flux. The
higher gravity suggests that \cfbdsA\ is older and higher mass than
\ugps, consistent with their relative tangential velocities
($47\pm5$~\kms\ and $19\pm3$~\kms, respectively). Of course, tangential
velocities are only a statistical proxy for age so this comparison
should be treated accordingly.

To assess the impact of component~B on the integrated-light spectrum, we
used the BT-Settl-2009 model atmospheres to create composite spectra. We
chose models with 600--700~K for the primary and 400--500~K for the
secondary, matching pairs of models with 200--250~K temperature
differences and the same surface gravity. The pairs of spectra were
scaled to agree with the observed $J$-band or $H$-band flux ratio and
summed to form artificial composites. We then added noise based on the
measured S/N of the VLT spectrum and measured the same indices. We found
that the blended composites had very similar index values as the input
primary spectrum alone, with systematic effects at the 5--20\% level,
typically 10\%. Such amplitudes are not sufficient to affect the
interpretation of Figure~\ref{fig:indices}.

\subsubsection{Model atmosphere fitting \label{sec:atmmodels}}

To further place \cfbds{AB} in context, we perform a homogeneous
comparison of the near-IR spectra of very late-T dwarfs to
low-temperature atmospheric models. We restrict our analysis to
$\ge$T8~dwarfs with parallaxes, excluding the aforementioned possible
binary T8~dwarf 2MASS~J0939$-$24, but also include two benchmark
T7.5~dwarfs (Table~\ref{table:atmmodels}).
We use the published $J$-band photometry to establish the absolute flux
calibration of the spectra, again assuming the primary dominates the
integrated-light spectrum. For \cfbdsAB, we again assumed that the
integrated-light spectrum is dominated by component~A and normalized it
using the resolved $J$-band photometry of component~A.

We fit three grids of atmospheric models: (1)~the Ames-Cond models
\citep{2001ApJ...556..357A} spanning $\Teff = 400-1200$~K
($\Delta\Teff=100$~K) and $\logg = 4.0-6.0$ ($\Delta\logg = 0.5$);
(2)~the Tucson models \citep[hereafter BSL03]{2003ApJ...596..587B},
which are irregularly gridded between $\Teff \approx 150-850$~K and
$\logg \approx 3.3-4.8$~(cgs); and (3) the aforementioned BT-Settl-2009
models spanning 400--1200~K and \logg~=4.0--6.0~dex ($\Delta\Teff=50$~K
above 600~K and 100~K below 600~K). All grids are for solar metallicity
compositions and chemical equilibrium.
The models were gaussian smoothed to the resolution of the data and
interpolated onto the wavelength grid of the observed spectrum.

The atmospheric models have known systematic errors, particularly in the
$Y$-band and the red side of the $H$-band as a result of imperfect
modeling of the pressure-broadened \ion{K}{1} doublet in the far-red and
the incomplete list of CH$_4$ lines, respectively. We therefore restrict
our fits to the 1.15--1.35~$\mu$m, 1.45--1.6~$\mu$m, and
1.95--2.25~$\mu$m spectral regions. The model emergent spectra are
scaled to the flux calibrated spectra by minimizing the $\chi^2$
statistic (e.g., Equation 5 and 6 of \citealt{2009ApJ...706.1114B}). For
observed spectra without measurement errors (CFBDS~J0059$-$01 and \ugps)
we perform a least-squares fit to the data (Equation~1 of
\citealt{stephens09-irac} with $w_i$=1, and the scaling factor is solved
for by equating the derivative of $G_1$ with zero).

The best-fitting models are summarized in Table~\ref{table:atmmodels}
and plotted in Figure~\ref{fig:modelfits}. 
The Ames-Cond models yield effective temperatures between 700--800~K and
a gravity of 4.5~dex for almost all objects; the temperatures for the
latest-type objects are $\approx$100--200~K hotter than analyses from
their discovery papers (which use variants of the BT-Settl models).
The BSL03 models typically produce much cooler temperatures (420--800~K)
and somewhat higher gravities (4.6--4.8~dex), with the results for the
earlier-type objects often at the edge of the model grid, indicating a
broader set of models is probably needed.
The BT-Settl-2009 model results are usually intermediate between those
from the other two models.
The fits show that CFBDS~J0059$-$01, \cfbds{AB}, and \ugps\ are the
three coldest objects in the sample, with the Ames-Cond models
indicating all have comparable temperatures while the BSL03 and
BT-Settl-2009 models indicate that \cfbds{AB} is the warmest. 
(Note that in terms of the absolute values derived from the fitting, the
analysis in the Appendix on the resulting spectroscopic distances
suggests that the temperatures may be somewhat overestimated.)

Again, to assess the effect of component~B on the analysis of the
integrated-light spectrum, we took the artificial noisy composite
spectra described in \S~\ref{sec:indices} and applied the same model
atmosphere fitting procedure. We found that the \Teff's determined for
the composite spectra were at worst 50~K colder than those of the input
primary spectrum alone, and often the same. The fitted surface gravities
were the same for the composites and the primary spectrum. Thus, the
blended light from the much fainter secondary appears to have little
consequence for our analysis, as expected.

We also examine another method to derive physical properties from
atmospheric models. Although the distance estimates using the scaling
factor $(R/d)^2$ may differ substantially from the true distances, the
known distance can be used in combination with the scaling factor, a
grid of radii, and a likely age range to infer a temperature and mass
range for the object. 
This is effectively using the scaling factor as a proxy for temperature,
and it has the benefit of relying on the absolute level of the model
emergent flux density rather than the detailed shape of the synthetic
spectrum, which may have systematic errors. Using the results from
Table~\ref{table:atmmodels}, we compute the range of effective
temperatures and masses for the three sets of models. The results are
shown in Table~\ref{table:tempmass} assuming an age range of 1--10~Gyrs,
except for Ross~458C where we assume an age of 0.1--1.0~Gyr
\citep{2010MNRAS.405.1140G}.
The BSL03 models are limited at high gravities by an upper mass cutoff
of 25~$M_\mathrm{Jup}$ so they only provide lower limits on
$T_\mathrm{eff}$ and mass. The resulting range of temperatures and
masses is similar to those estimated in previous work
\citep{2006liu-hd3651b, 2008MNRAS.391..320B, 2008A&A...482..961D}
This analysis indicates that \cfbds{A} has a similar effective
temperature (540--660~K) and mass (12-42~M$_\mathrm{Jup}$) to the T9
dwarfs ULAS~J0034$-$00 and ULAS~J1335+11, with CFBDS~J0059$-$01 and
UGPS~J0722$-$05 being perhaps slightly cooler, in accord with the
relative ordering of their absolute magnitudes.

% - - - - - - - - - - - - - - - - - - - - - - - - - - - %

\subsection{Resolved Colors and Magnitudes\label{sec:phot}}

In accord with the integrated-light spectrum, the near-IR colors of
component~A are similar to those of the known T8--T10 dwarfs, with its
$(H-K)$ color of $-0.45\pm0.26$~mag being somewhat bluer than most
objects, though the uncertainty is large.
As discussed above, the bluer color is likely due to a higher surface
gravity (\ie, an older age) compared to most other late-T~dwarfs.

Our near-IR parallax for the system provides the absolute magnitudes for
component~B and reveals its extreme nature. Compared to the T10~dwarf
\ugps, \cfbds{B} is fainter by 1.4$\pm$0.4, 1.9$\pm$0.3, and
2.0$\pm$0.4~mag in its $JHK$ absolute magnitudes, respectively.
Being significantly fainter than previously known brown dwarfs
(Figure~\ref{fig:cmd}), the properties of \cfbds{B} provide a first look
at the empirical trends beyond spectral type T10, especially
enlightening given the uncertainties in model atmosphere predictions for
such cold objects \citep[e.g.][]{2003ApJ...596..587B}.
The very blue $(CH_4s-H)$ color indicates strong \meth\ absorption
persists at $H$-band, comparable to the known T8--T9 dwarfs which have a
$(CH_4s-H) \approx -0.5$ to $-$0.6~mag (\eg, Figure~2 of
\citealp[][]{2005AJ....130.2326T} and Section~3.2 of
\citealp{liu08-2m1534orbit}).

The near-IR colors of \cfbds{B} are very blue
(Figure~\ref{fig:properties}), with its $H-K$ color is similar and the
$J-K$ color slightly bluer than previously known T~dwarfs (except for the
T6p dwarf 2MASS~0937+29, which is suspected to have low metallicity;
\citealp{burg01}).
For its absolute magnitude, its $J-K$ color is much redder than those
predicted by the Lyon/Cond \citep{2003A&A...402..701B} and
\citet[][hereafter SM08]{2008ApJ...689.1327S} models, and to a lesser
degree the BSL03 models as well.
The most intriguing behavior is seen in $J-H$, where the blueward trend
with spectral type among the T~dwarfs appears to reach an inflection
point with the T8--T10 dwarfs, and then \cfbds{B} continues to bluer
colors.

% - - - - - - - - - - - - - - - - - - - - %

\subsection{Physical Properties from Evolutionary Models \label{sec:properties}}

How much cooler is \cfbds{B} than previously known late-T dwarfs? Using
its absolute magnitudes as a simple proxy for \Lbol\ (\ie, assuming the
same bolometric corrections as other T9--T10 objects), its \Lbol\ is a
factor of $\approx$4--6$\times$ smaller. Old brown dwarfs have
essentially constant radii, so this would correspond to an object that
is a factor of $\approx$1.5 cooler. The T9--T10 objects have estimated
\Teff\ of 500--600~K from model atmosphere fitting and their bolometric
luminosties \citep{2009ApJ...695.1517L, 2010ApJ...720..252L,
  2010A&A...524A..38M, 2010MNRAS.408L..56L}; simple scaling would then
make \cfbds{B} $\approx$350-400~K.

With \Lbol\ estimates for the individual components and an assumed age,
evolutionary models can fully determine the remaining physical
properties of the two components \citep[e.g.][]{2000ApJ...541..374S}.
This approach is desireable as predictions of \Lbol\ are less dependent
on the assumed model atmospheres than predictions of the absolute
magnitudes \citep{2000ApJ...542..464C, 2008ApJ...689.1327S}. This
approach has also been widely used for temperature determinations for
field L~and T~dwarfs \citep[e.g.][]{gol04}. Of course, such estimates
should be treated with caution, since \cfbdsB\ resides in a
low-temperature, low-luminosity regime where the models have not been
tested against any observations. (In the case of L~dwarfs,
\citealp{2008arXiv0807.2450D} have found that for a given luminosity and
assumed age, the evolutionary models may overestimate the masses.)

To estimate \Lbol, we adopt $J$-band bolometric corrections from the
coolest T~dwarfs with parallaxes: ULAS~J0034$-$00 (T9; $BC_J =
1.96\pm0.11$~mag), CFBDS~J0059$-$01 (T9; $1.87\pm0.21$~mag), and
ULAS~J1335+11 (T9; $1.82\pm0.09$~mag) from \citet{2010A&A...524A..38M},
and \ugps\ (T10; $1.36\pm0.21$~mag) from \citet{2010MNRAS.408L..56L}.
The unweighted average is $1.75\pm0.27$~mag, and we apply this to both
components of \cfbdsAB.
The resulting values for \Lbol/\Lsun\ are $(1.1\pm0.4) \times 10^{-6}$
and $(2.0\pm0.9) \times 10^{-7}$
for components~A and~B, respectively. In comparison, CFBDS~J0059$-$0114
and \ugps\ have \Lbol/\Lsun = $(7.2\pm1.2) \times 10^{-7}$ and
$(9.2\pm3.1) \times 10^{-7}$, respectively, making \cfbds{B} a factor of
$\approx$4--5 less luminous than these two objects.

Table~\ref{table:binary} presents the derived physical quantities for
ages of 1.0 and 5.0 Gyr. The \Lbol\ uncertainty is propagated into the
calculations in a Monte Carlo fashion. The inferred temperature of
component~B is exceptionally low, 360--380~K from the Lyon/COND models
and 350--380~K from the \citet{1997ApJ...491..856B} ones. Including the
random uncertainties from the Monte Carlo results, we adopt a final
\Teff\ estimate of 370$\pm$40~K. The model-derived mass is 6--14 (7--17)
for the Lyon/COND (Burrows) models, possibly placing the object below
the notional 13~\Mjup\ dividing line between planets and brown dwarfs
based on the deuterium-burning limit. Overall, the physical properties
of \cfbdsB\ overlap those of the gas-giant planets found from radial
velocity surveys. In fact, both components of \cfbds{AB} would reside in
the planetary-mass regime and have retained their initial deuterium
abundance for ages of $\lesssim$1~Gyr.

To gauge the systematic effects, we also compute physical properties
using the BSL03 models, which provide predictions for absolute
magnitudes. (The Lyon/COND models also provide absolute magnitudes, but
these disagree more strongly with the color-magnitude data in
Figure~\ref{fig:cmd}.) We compare the BSL03 models to the $J$-band
observations, as this bandpass represents the peak of the near-IR SED.
(Note that this is largely equivalent to using the BSL03 atmospheric
models for their bolometric corrections.) The results are listed in
Table~\ref{table:binary} and agree reasonably well with those derived
from \Lbol\ and assuming the bolometric corrections from the known
T9--T10 dwarfs.

% - - - - - - - - - - - - - - - %
%       Deuterium burning       %
% - - - - - - - - - - - - - - - %
Measuring the differential deuterium abundance would help constrain the
age of the system, analogous to the binary lithium test proposed by
\citet{2005astro.ph..8082L}.
The deuterium depletion timescales are long for such low-mass objects,
so the exact age at which they become fully depleted is fuzzy. Using a
finely interpolated version of the Lyon/COND models
\citep{2003A&A...402..701B} and constrained by the observed luminosity
of each component, we find that the full range of 99\% to 0\% deuterium
content for the primary is 0.5--1.3~Gyr, while the the 99--0\% deuterium
range for the secondary is 2.3--5.4~Gyr. Thus, for ages of $<$0.5~Gyr,
both components retain their deuterium and would have masses of $<$9 and
$<$4~\Mjup.
Between 0.5--2~Gyr, the primary is depleted but the secondary retains
its deuterium. And for ages older than 2~Gyr both components are
depleted and are $>$20 and $>$10~\Mjup. (\citealp{allers09-sdss2249}
report the discovery of a young field L~dwarf binary whose estimated
luminosities and age indicate that the two components may be in the
process of differential deuterium depletion.)
Note that the deuterium abundance is expected to be difficult to measure
\citep{2000ApJ...542L.119C}. An additional complication is the
theoretical predictions for the D-burning limit depend on the
metallicity, initial D and He abundances, and atmospheric boundary
conditions and can range from $\approx$11--14~\Mjup\
\citep{2010arXiv1008.5150S}.

% - - - - - - - - - - - - - - - - - - - - %

\subsection{Orbital Period}

To estimate the orbital period of the binary, we adopt the statistical
conversion factor between projected separation and true semi-major axis
from \citet{dupuy2011-eccentricity}. They offer several choices, based
on the underlying eccentricity distribution and degree of completeness
to finding binaries by imaging (``discovery bias''). We adopt the
eccentricity distribution from their compilation of ultracool visual
binaries with high quality orbits and assume their case of modest
discovery bias, namely that the smallest possible projected separation
for detection is half the semi-major axis. This gives a multiplicative
correction factor of 1.07$^{+0.42}_{-0.27}$ (68.3\% confidence limits)
for converting the projected separation into semi-major axis (compared
to $1.10^{+0.92}_{-0.35}$ for a uniform eccentricity distribution with
no discovery bias).
The resulting orbital period estimates range from 20--35~years for ages
of 1.0--5.0 Gyr, albeit with significant uncertainties
(Table~\ref{table:binary}). Dynamical masses for ultracool binaries may
be realized from orbital monitoring covering $\gtrsim$30\% of the period
\citep[e.g.][]{2004A&A...423..341B, liu08-2m1534orbit,
  2009ApJ...706..328D}. Thus, \cfbdsAB\ warrants continued high angular
resolution imaging to determine its visual orbit, which can then be
combined with the parallax to directly measure its total mass.

%----------------------------------------------------------------------%

\section{Discussion: Y or Y not?}

Definition of the Y~spectral type awaits discovery of objects whose
spectra are unambiguously different from the T dwarfs, in the same
fashion as the distinction between M~and L~dwarfs and between L~and
T~dwarfs.
The near-IR spectral progression observed for the T8--T10 dwarfs is
subtle, as the \meth\ bands are nearly fully absorbed. Such late-type
objects are distinguished from each other by only modest changes in the
shape of the near-IR flux peaks \citep[e.g.][]{2008A&A...482..961D,
  2008MNRAS.391..320B, 2010MNRAS.408L..56L}, though the thermal-IR
($\lambda>3$~\micron) colors are more distinctive
\citep{2010ApJ...710.1627L}.

Recent work has highlighted near-IR spectral features that might arise
from \ammonia\ \citep{2006astro.ph.11062S, 2008A&A...482..961D,
  2010MNRAS.408L..56L} but not in a definitive fashion. 
Model atmospheres suggest the onset of \ammonia\ at $\approx$600~K but
also indicate the effect could be indistinctive
\citep{2003ApJ...596..587B, 2007ApJ...667..537L}. The emergence of
narrow intra-band features, possibly linked to \ammonia\ absorption, is
difficult to disentangle from the effects of differing gravities and
metallicities. Also, the available opacities for \meth\ and \htwoo\ are
known to be incomplete, complicating the identification of new species.
In short, the current census does not seem to compel the use of the
Y~spectral type. (Of course, future identification of unambiguous
examples may lead to re-classification of some known objects as early
Y~dwarfs.)

Overall, \cfbds{B} is the most promising candidate to date for the
hypothesized Y~spectral class, given its extremely low luminosity,
atypical near-IR colors, and cold inferred temperature.
While spectroscopy is needed to place it in full context with the known
T~dwarfs, the singular nature of this object is unambiguous, thanks to
our parallax measurement.
CFBDSIR~J1458+1013B is $\approx$1.5--2.0~magnitudes less luminous in the
near-IR than any previously known brown dwarf and therefore is cooler
than any known brown dwarf.\footnote{From a deep {\sl Spitzer}
  extragalactic survey, \citet{2010AJ....139.2455E} have identified a
  very faint, low/no proper motion source whose exceptionally red
  thermal-IR colors and non-detection in the near-IR suggest a
  $\lesssim$450~K object at $\gtrsim$70~pc. Direct comparison with
  \cfbds{B} is not possible, since there are no photometric detections
  in common between the two objects. Their object's $H$-band
  non-detection ($H>24.2$~mag) is consistent with \cfbds{B}, which would
  be $H=24.9$~mag at 70~pc.} For reference, two magnitudes in absolute
magnitude spans most of the T~dwarf sequence, from about T0 to T6
\citep[e.g.][]{2006astro.ph..5037L, 2010A&A...524A..38M}. On this basis
alone, this object is a significant leap. For comparison, the T10~dwarf
\ugps\ is only $\approx$0.2~mag fainter than CFBDS~J0059$-$01 in
absolute magnitude, and among the small sample of $>$T8 dwarfs, the
differences in absolute magnitudes are $\lesssim$0.5~mag between
objects.
For this reason, a spectral type estimate of $>$T10 seems appropriate
for \cfbds{B}, as opposed to $\ge$T10. (A naive extrapolation of the
linear trend in absolute magnitude with T~subclass would make the object
around T12.) 
While it could be that near-IR spectra changes so slowly with
temperature that a broad luminosity range of objects all appear to be
the same as the \ugps, this seems unlikely given the behavior of the
hotter T~subclasses.

Regardless of its final spectral classification, \cfbdsB\ provides a
gateway into new processes in low-temperature atmospheres. Models by
\citet{2003ApJ...596..587B} predict the disappearance of the very broad
sodium (0.59~\micron) and potassium (0.77~\micron) resonance lines at
$\Teff\approx450-500$~K, comparable to the estimated temperature of
\cfbds{B}. (See also \citealp{2007ApJ...667..537L}.) This change might
be detected from resolved far-red optical spectroscopy and photometry
from \HST. Also, Burrows \etal\ predict the formation of photospheric
water clouds at $\approx$400--500~K (with hotter temperatures for higher
surface gravities), though this phenomenon has little impact on their
model emergent spectra. At $\approx$300--400~K, they predict the
cessation of the blueward trend in near-IR colors with decreasing
temperature. Such signatures might be delineated with resolved
multi-band far-red optical and near-IR data of \cfbds{AB}.
Our current measurements indicate that the trends of near-IR colors with
spectral type continue without enormous changes, but that they may be
non-monotonic. Indeed, the subtle changes seen in the near-IR SEDs of
late-T dwarfs, combined with the strong alkali line depletion predicted
at $\lesssim$500~K, suggest that far-red optical wavelengths may become
valuable for spectral classification of the coolest new discoveries.

The fact that \cfbds{B} resides in a binary system of known distance
offers additional future opportunties for understanding the coldest
brown dwarfs. The estimated mass ratio ($q\approx0.5$) is uncommonly
small compared to most field systems, representing only $\lesssim$5\% of
known binaries \citep{2010liu-2m1209}. Measuring accurate \Lbol\ and
\Teff\ for the two components could test the joint accuracy of
evolutionary and atmospheric models under the constraint of coevality
\citep[e.g.][]{1999ApJ...520..811W, 2009ApJ...704..531K,
  2010liu-2m1209}. Furthermore, given their identical composition and
similar surface gravities, comparing the SEDs of the two components
directly probes the influence of decreasing effective temperature,
without other physical parameters to consider. So the binary may yield
valuable insights about the transition from T~dwarfs to Y~dwarfs.
Finally, the estimated orbital period of $\approx$20--30~years means a
direct dynamical mass estimate is feasible over the next several years;
such a mass benchmark would provide a stringent test of substellar
models. A dynamical mass would also help to demarcate the low-mass limit
for star formation.

%----------------------------------------------------------------------%

\acknowledgments

It is a pleasure to thank Jim Lyke, Gary Punawai, Cynthia Wilburn, and
the Keck Observatory staff for assistance with the LGS AO observing and
the staffs of CFHT and VLT for carrying out the queue imaging and
spectroscopy.
% This work has benefitted from discussions with \TBD. 
We thank Sandy Leggett and Didier Saumon for making published results
readily available and thank Kathleen Brandt for assistance while this
paper was being written. Our research has employed the 2MASS data
products; NASA's Astrophysical Data System; the SIMBAD database operated
at CDS, Strasbourg, France; and the SpeX Prism Spectral Libraries
maintained by Adam Burgasser at http://www.browndwarfs.org/spexprism.
MCL, TJD, and BPB acknowledge support for this work from NSF grants
AST-0507833 and AST-0909222. Support for this work to TJD was partially
provided by NASA through Hubble Fellowship grant
\hbox{{\#}HF-51271.01-A} awarded by the Space Telescope Science
Institute, which is operated by the Association of Universities for
Research in Astronomy, Inc., for NASA, under contract NAS 5-26555.
Finally, the authors wish to recognize and acknowledge the very
significant cultural role and reverence that the summit of Mauna Kea has
always had within the indigenous Hawaiian community.  We are most
fortunate to have the opportunity to conduct observations from this
mountain.

{\it Facilities:} \facility{Keck II Telescope (LGS AO, NIRC2),
  Canada-France-Hawaii Telescope (WIRCAM), VLT-UT2 (X-Shooter)}

%----------------------------------------------------------------------%



\appendix

\section{Spectroscopic Distance Estimates for Late-T Dwarfs}

Using the fitting results from \S~\ref{sec:atmmodels}, we can test the
accuracy of atmospheric models by comparing the measured distances to
model-derived spectroscopic distances following the method of
\citet{2009ApJ...706.1114B}. The scaling factor between the model
emergent spectra and the observed spectra is equal to $(R/d)^2$, so if a
radius is known (or assumed) then the distance can be computed. Bowler
\etal\ estimated distances to four late-M and L subdwarfs using this
method and found agreement to within $\sim$10\% of published parallaxes.
Several subsequent studies have used this technique to estimate
distances for late-T dwarfs without parallaxes
\citep[e.g.][]{2011ApJ...726...30M}, but this method has yet to be
tested for such objects. We do so now with the Ames-Cond, BSL03, and
BT-Settl-2009 models.

To compute model-derived spectroscopic distances, for the Ames-Cond and
BT-Settl-2009 atmospheric models we use radii from the companion grid of
Lyon/Cond evolutionary models \citep{2003A&A...402..701B} interpolated
across ages, effective temperatures, and surface gravities. For the
BSL03 models, we use a power law fit for the radius as a function of
\Teff\ and \logg\ from the \citet{1997ApJ...491..856B} evolutionary
models (Equation~1 of BSL03). We compute errors in the spectroscopic
distances from two components. (1)~Measurement uncertainties from the
finite S/N of the spectra and the absolute flux calibration from the
$J$-band photometry are accounted for in a Monte Carlo fashion. We adopt
the median and RMS of the resulting distributions as one version of the
results.
(2) In addition, we account for systematic uncertainties in the models
by assuming normally distributed errors of 50 K and 0.25 dex in the
fitted \Teff\ and \logg. The standard deviation of the resulting
spectroscopic distance distribution is reported along with the
uncertainty due to random errors.

Table~\ref{table:atmmodels} summarizes the results, with both sets of
uncertainty calculations. The agreement between the spectroscopic
distances and the parallactic distance range from $\approx$10\% to about
a factor of two, with no apparent trend in accuracy with distance,
spectral type, or model grid. The large scatter likely arises from the
sensitivity of the distance estimate to the best-fitting model \Teff. 
The typically larger spectroscopic distances compared to parallactic
distances imply an over-estimate of \Teff\ in the best-fitting model,
because the scaling factor is highly sensitive to temperature (absolute
flux level) and very weakly dependent on gravity.

%%======================================================================%%




%\begin{thebibliography}{}
%\end{thebibliography}

%\bibliography{/Users/mliu/tex/bibtex/mliu}
%\bibliographystyle{apj}

%======================================================================%
\clearpage

\begin{figure}
%\vskip -1in
\hskip -0.5in
\hbox{
\includegraphics[height=2.2in,angle=0]{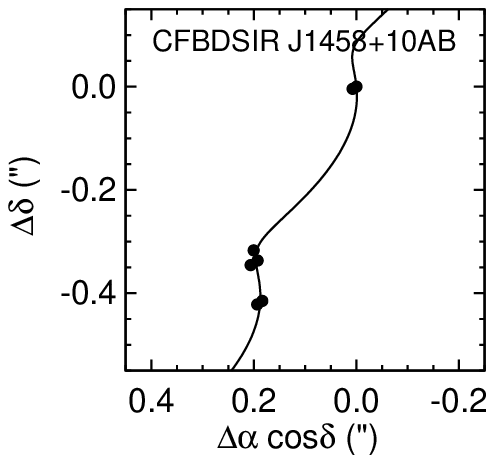}	
%\includegraphics[height=2.2in,angle=0]{CFBDS1458+10_ra-de.ps}	
%\hskip -0.5in
\includegraphics[height=2.2in,angle=0]{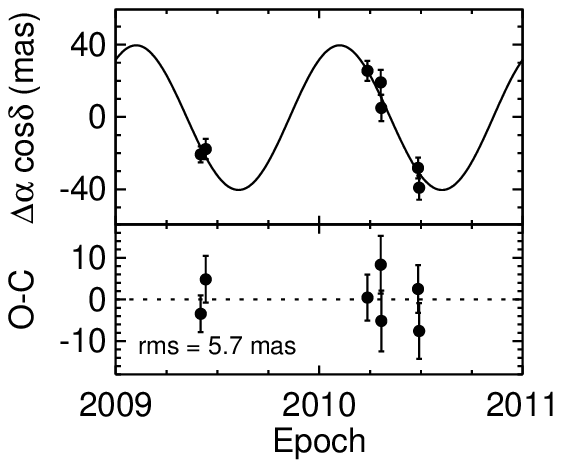}	
%\includegraphics[height=2.2in,angle=0]{CFBDS1458+10_ra-time.ps}	
%\hskip -0.7in
\includegraphics[height=2.2in,angle=0]{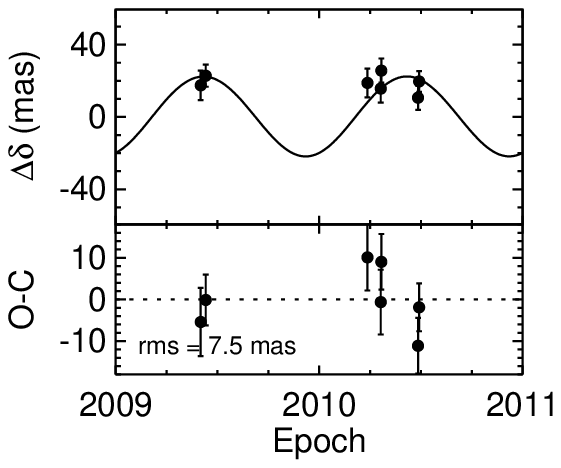}
}
\vskip 1ex
\caption{\normalsize {\em Left:} Relative astrometry from CFHT/WIRCam
  $J$-band imaging of \cfbds{AB} that shows both proper motion and
  parallax. The origin corresponds to the first epoch of data.  {\em
    Middle and Right:} Top panels show relative astrometry in $\alpha$
  and $\delta$, respectively, as a function of Julian year after
  subtracting the best-fit proper motion. (This is for display purposes
  only; in our analysis we fit for both proper motion and parallax
  simultaneously.) Bottom panels show the residuals after subtracting
  both parallax and proper motion. \label{fig:parallax}}
\end{figure}

\begin{figure}
%\vskip -1in
%\hskip -0.5in
%\includegraphics[width=3in,angle=270]{1458_UGPS0722_YJHK_Jnorm.ps}
%\includegraphics[width=5in,angle=90]{1458_UGPS0722_YJHK_Jnorm_YJ+H_overlays.pptx.ps}
%\centerline{\includegraphics[width=5.5in,angle=0]{CFBDIR1458_UGPS0722_comp.eps}}
\centerline{\includegraphics[width=7in,angle=0]{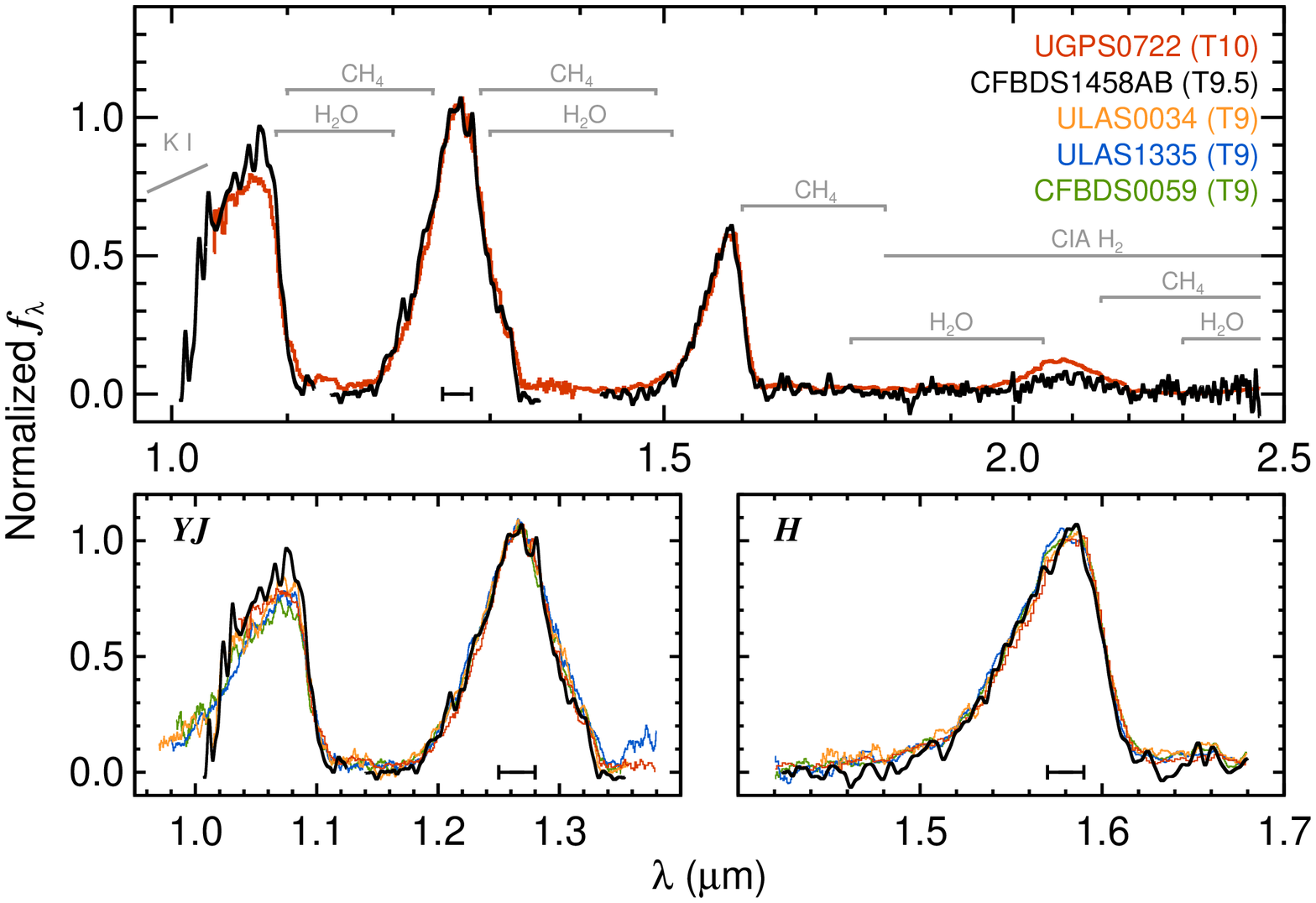}}
%\centerline{\includegraphics[width=7in,angle=0]{cfbdir1458_latet_comp_3panel.eps}}
%\raise -3in
%\vbox{
%\includegraphics[width=3in,angle=0]{CFBDIR1458_compfig.eps}}
\vskip -3ex
\caption{\normalsize The integrated-light near-IR spectrum of \cfbdsAB\
  (black line) compared with other late-T dwarfs. The spectra have been
  smoothed for clarity. The upper panel shows a comparison with the
  T10~dwarf \ugps\ (red line; \citealp{2010MNRAS.408L..56L}). The lower
  panels show the $YJ$ (left) and $H$-band (right) comparison with
  \ugps\ and three T9~dwarfs \citep{2007MNRAS.381.1400W,
    2008MNRAS.391..320B, 2008A&A...482..961D}. The black horizontal bars
  near the bottom of the plots show the wavelength ranges used to
  normalize the spectra (1.25--1.28~\micron\ for the top and bottom left
  plots, and 1.57--1.59~\micron\ for the bottom right plot).  Note that
  for the upper panel, the x-axis uses a log scale to help make the
  bluer bandpasses more visible.   
  \label{fig:spectra}}
%(median 17, smooth5) ISAAC reso=500, 6 pix per spectral resolution element. 
\end{figure}

% \begin{figure}
% \begin{tabular}{cc}
% \includegraphics[width=6.0cm,angle=270]{1458_UGPS07_0059_YJ.ps} &
% \includegraphics[width=6.0cm,angle=270]{1458_UGPS07_0059_H.ps} 
% \end{tabular}
% \caption{{$J$ (left) and $H$-band (right) spectra of the 3 coolest brown
%     dwarfs.CFBDS0059 is in dash-dotted red, UGPS0722 is in dashed cyan
%     and CFBDSIR1458 is in black line.}
% \label{fig:compare}
% \end{figure}

\begin{figure}
%\vskip -1in
\centerline{\includegraphics[width=5.5in,angle=0]{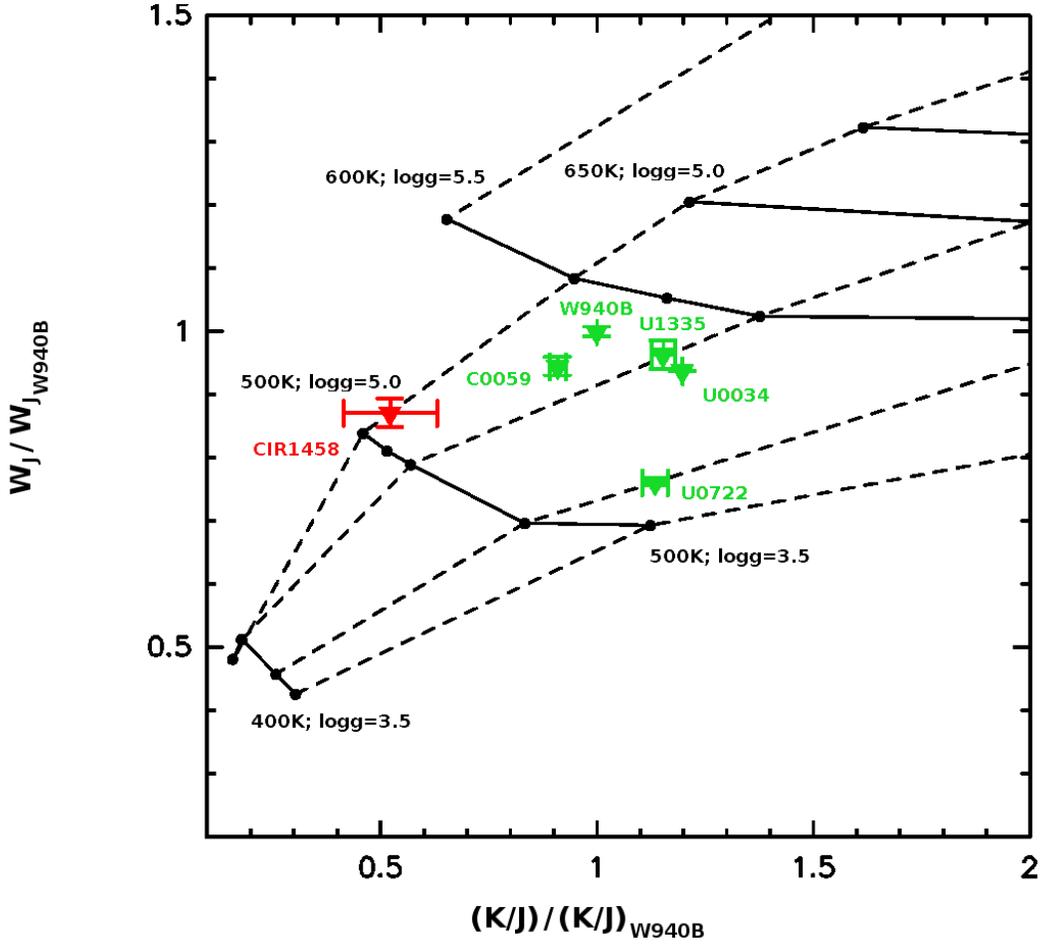}}
%\centerline{\includegraphics[width=5.5in,angle=0]{wj_jk_1458_weighted_mod.ps}}
%\vskip 1ex
\caption{\normalsize The $W_J$ and $J/K$ spectral indices for very
  late-T dwarfs compared to the solar-metallicity BT-Settl-2009 model
  atmospheres. (See Table~\ref{table:indices} for measurements and
  references.) The model grid has been scaled to agree with the
  T8.5~benchmark dwarf Wolf~940B ($\Teff = 575$~K, $\logg = 4.75$;
  \citealp{burningham08-T8.5-benchmark}). The plotted errors represent
  the 1$\sigma$ uncertainties in the index
  measurements.\label{fig:indices}}
\end{figure}

\begin{figure}
\hskip 0.25in
\includegraphics[width=5.5in,angle=0]{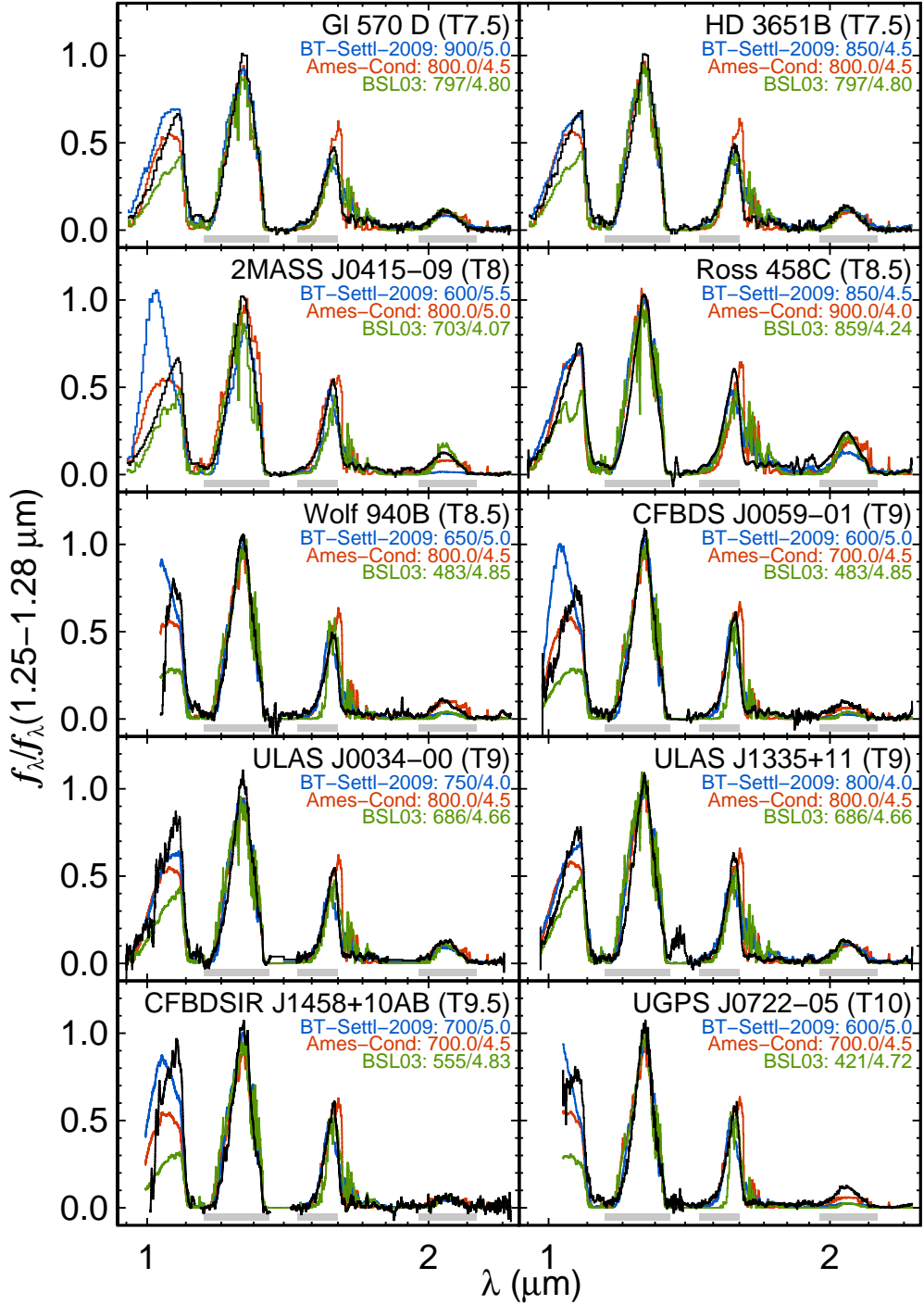}
\caption{\small Best-fitting BT-Settl-2009 (blue),
    Ames-Cond (red), and BSL03 (green) model atmospheres to late-T
    dwarfs with parallaxes. (See also Table~\ref{table:atmmodels}.) The
    gray horizontal bars show the wavelengths used in the fit, chosen to
    avoid known systematic errors in the models (e.g., the $Y$-band and
    the red side of the $H$-band) and regions of low S/N. The observed
    spectra are normalized to unity flux between 1.25--1.28~\micron, and
    the normalization of the models is an outcome of the fitting
    process. %, and the models are
    % scaled to minimize the $\chi^2$ value or rms deviation, depending
    % whether errors are available for the data (see $\S$XX).
    The observed spectra for Ross~458C, Wolf~940B, CFBDS~J0059$-$01,
    ULAS~J0034$-$00, and ULAS~J1335+11 are smoothed for clarity here,
    but the models were fit to the original (unsmoothed) data. Note that
    the x-axis uses a log scale, to help make the bluer bandpasses more
    visible.
    \label{fig:modelfits}}
\end{figure}

%\begin{figure}
%\includegraphics[width=8cm]{1458_H_model_500_400_5.0.ps}
%\caption{black:CFBDSIR1458, red, 400K, cyan 500K models
%\label{}}
%\end{figure}

%\begin{figure}
%\includegraphics[width=4in,angle=270]{1458_JHK_model_500_5.0.ps}
%\caption{black:CFBDSIR1458, red 500K model  {\color{red} ***KEEP THIS, OR TOSS IT?***}
%\label{fig:modelatm}}
%\end{figure}

\begin{landscape}
\begin{figure}
%%--------------------------------------------------%
%% submitted version
%%--------------------------------------------------%
% \vskip -0.5in
% \hskip 1in
% \includegraphics[width=4in,angle=0]{plot-cmd-H_JH.eps}
% \vskip -0.3in
% \hskip 1in
% \includegraphics[width=4in,angle=0]{plot-cmd-H_HK.eps}
% \vskip -0.3in
% \hskip 1in
% \includegraphics[width=4in,angle=0]{plot-cmd-H_JK.eps}
%
%%--------------------------------------------------%
%% revised version
%%--------------------------------------------------%
%\hskip -0.8in
%\hbox{
%\includegraphics[width=3.5in,angle=0]{plot-cmd-H_JK.eps}
%\hskip -1.2in
%\includegraphics[width=3.5in,angle=0]{plot-cmd-H_JH.eps}
%\hskip -1.2in
%\includegraphics[width=3.5in,angle=0]{plot-cmd-H_HK.eps}
%}
%--------------------------------------------------%
% revised version, changed to landscape
%--------------------------------------------------%
\hskip -1.1in
\hbox{
\includegraphics[width=4.45in,angle=0]{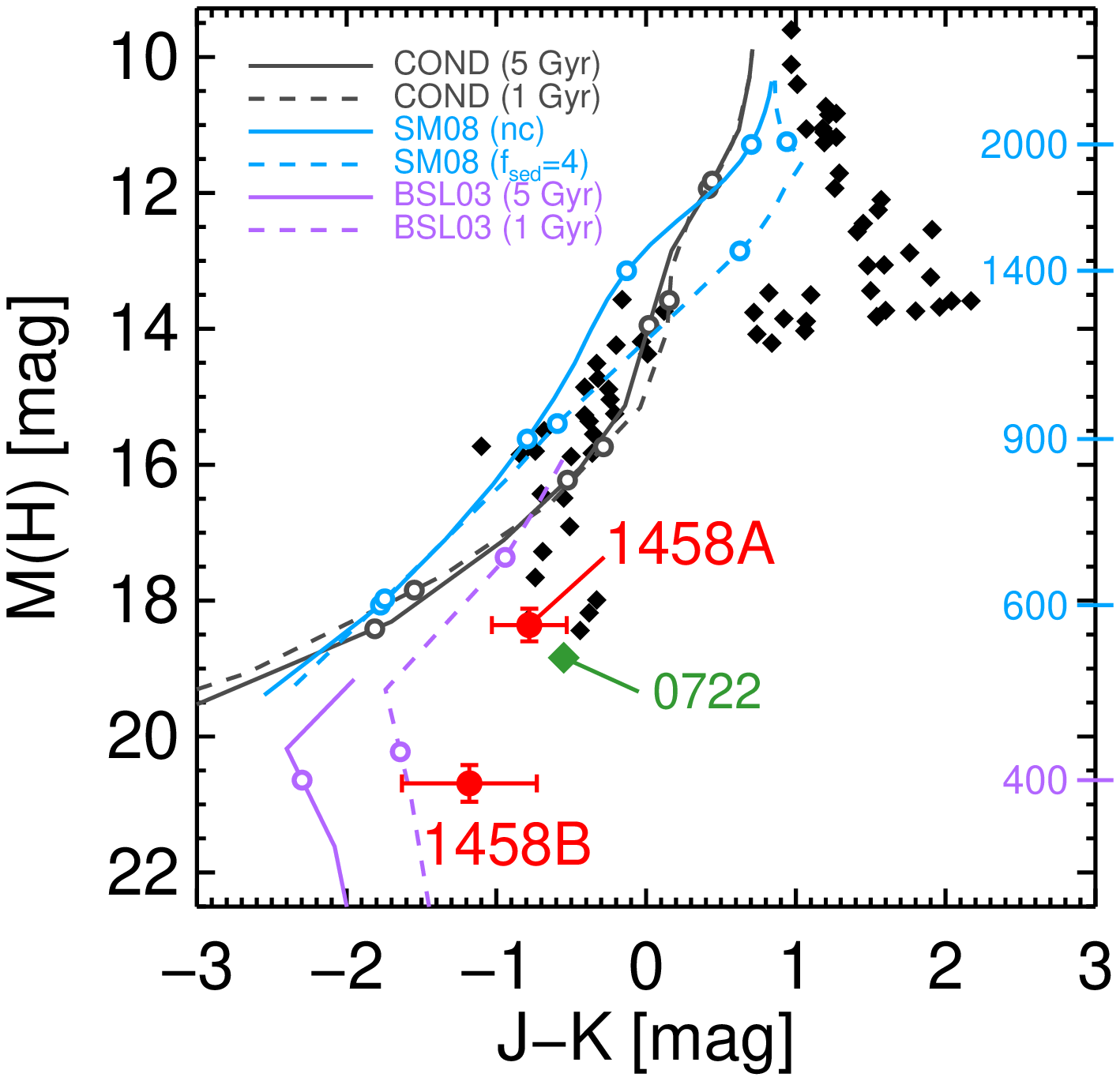}
\hskip -1.5in
\includegraphics[width=4.45in,angle=0]{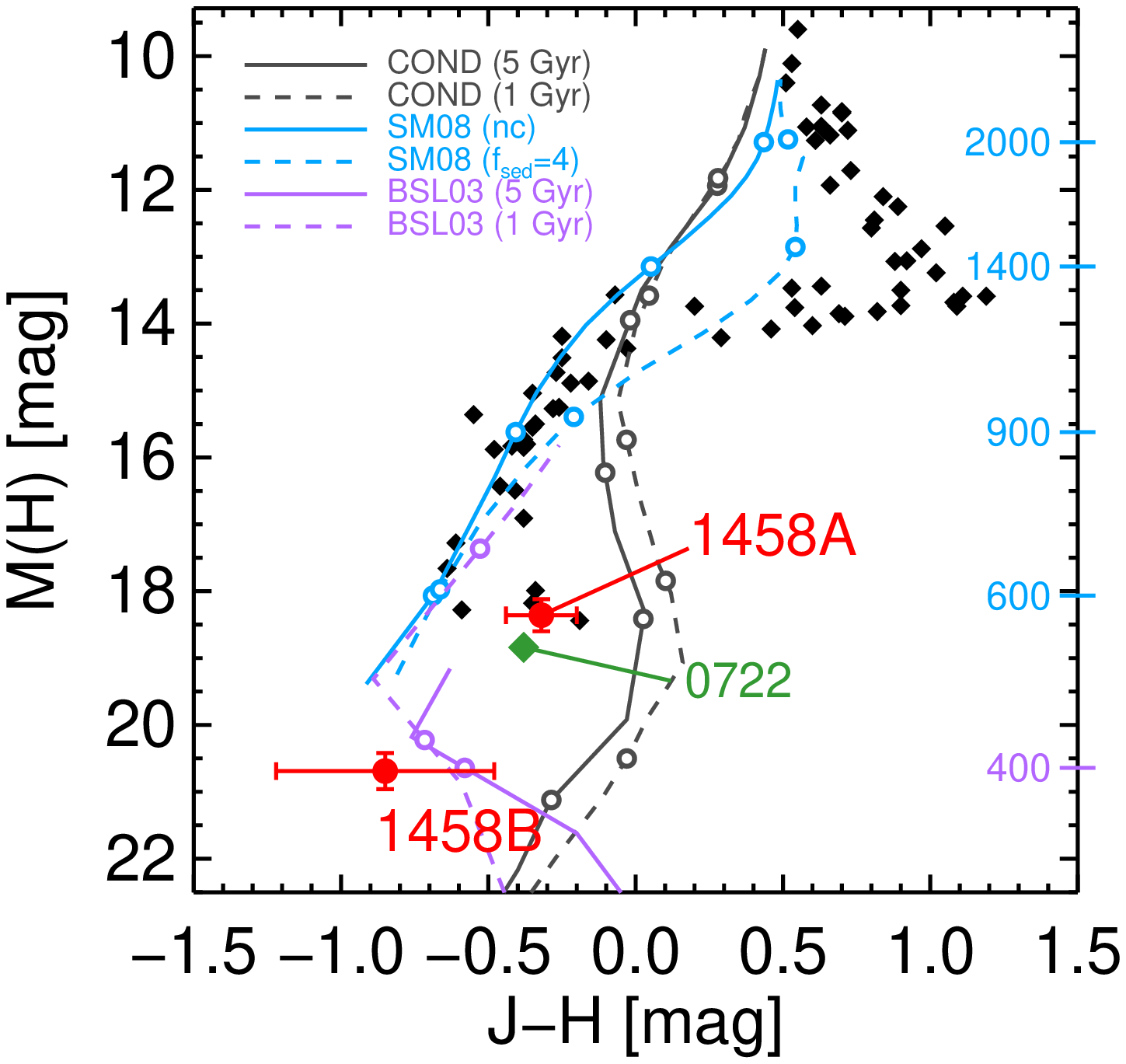}
\hskip -1.5in
\includegraphics[width=4.45in,angle=0]{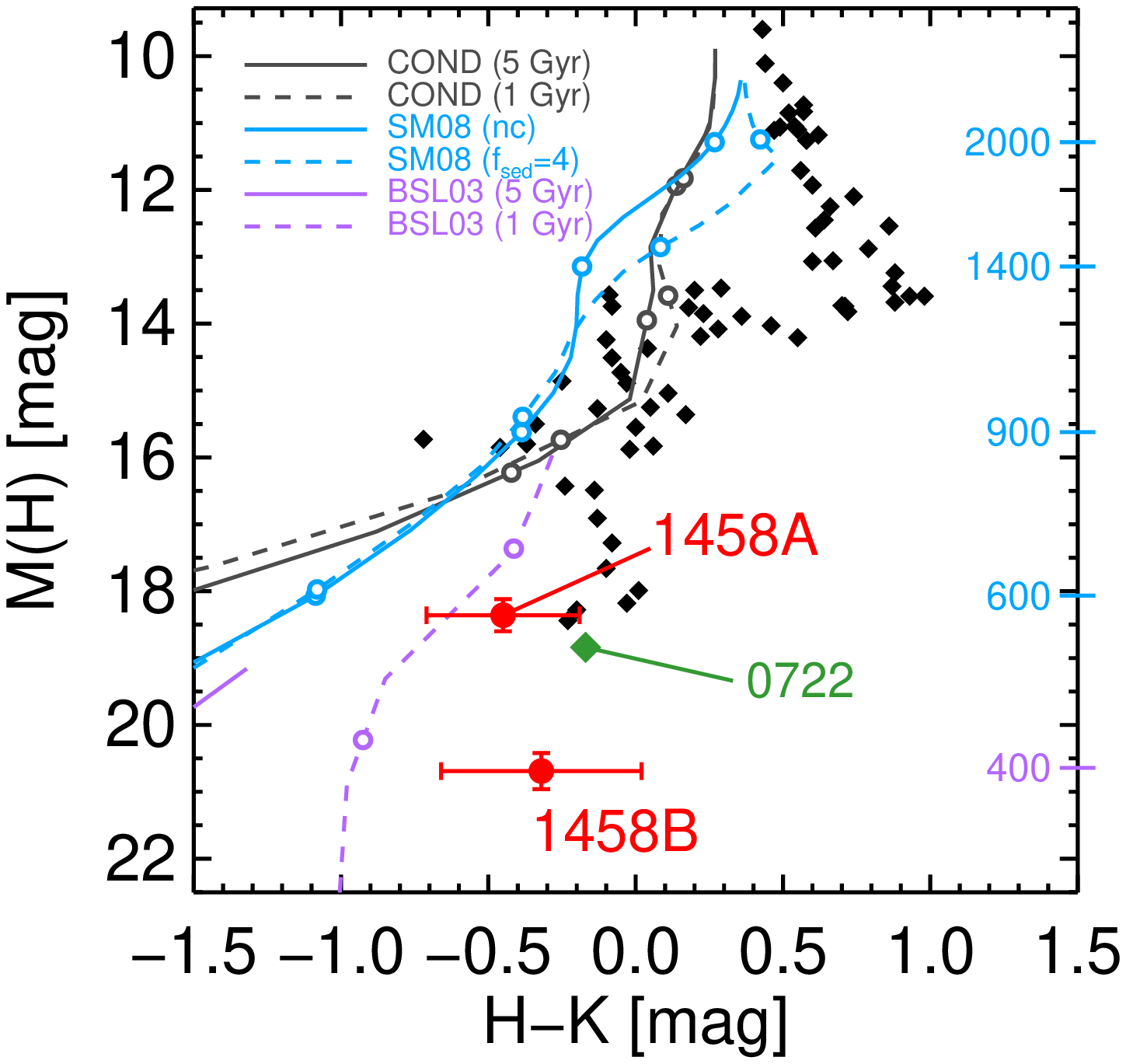}
}
%\vskip 1ex
\caption{\normalsize Color-magnitude diagrams comparing \cfbdsAB\ (red
  dots) to
  field L~and T~dwarfs with parallaxes \citep{2002AJ....124.1170D,
    2003AJ....126..975T, 2004AJ....127.2948V, 2010A&A...524A..38M},
  based on the compilation by \citet{2010ApJ...710.1627L} and plotted as
  filled diamonds. Known  
  binaries are excluded. The T10~dwarf \ugps\ is labeled. We also show
  the Lyon/COND \citep{2003A&A...402..701B},
  \citet{2008ApJ...689.1327S}, and \citet{2003ApJ...596..587B}
  models. 
  For the SM08 models, we plot the $\logg=5$ models that have a
  moderate cloud layer ($f_{sed}=4$) and those that have no clouds
  (``nc''). %For the SM08 models, we plot the cloudless (``nc'') models and those
  % with a moderate cloud layer ($f_{sed}=4$; see their paper for
  % details), both for $\logg = 5.0$.
  %
  The open circles on the model tracks demarcate a common set of
  \Teff's; for models plotted with a solid line, the corresponding
  \Teff\ values are listed on the right hand axis for the Saumon \&
  Marley models (which go down to 500~K) and then the BSL03 models for
  cooler temperatures. This combination is chosen for the labels
  because it agrees better with the data than the Lyon/COND models,
  though in general no single model agrees well with the loci of the
  latest T~dwarfs and \cfbds{B}. 
  %(Note that the model tracks do not
  %change significantly for different ages for the Lyon/COND and BSL03
  %models or for different surface gravities in the case of the SM08
  %models.)
  \label{fig:cmd}}
\end{figure}
\end{landscape}

\begin{figure}
\hskip -0.25in
\includegraphics[width=4in,angle=0]{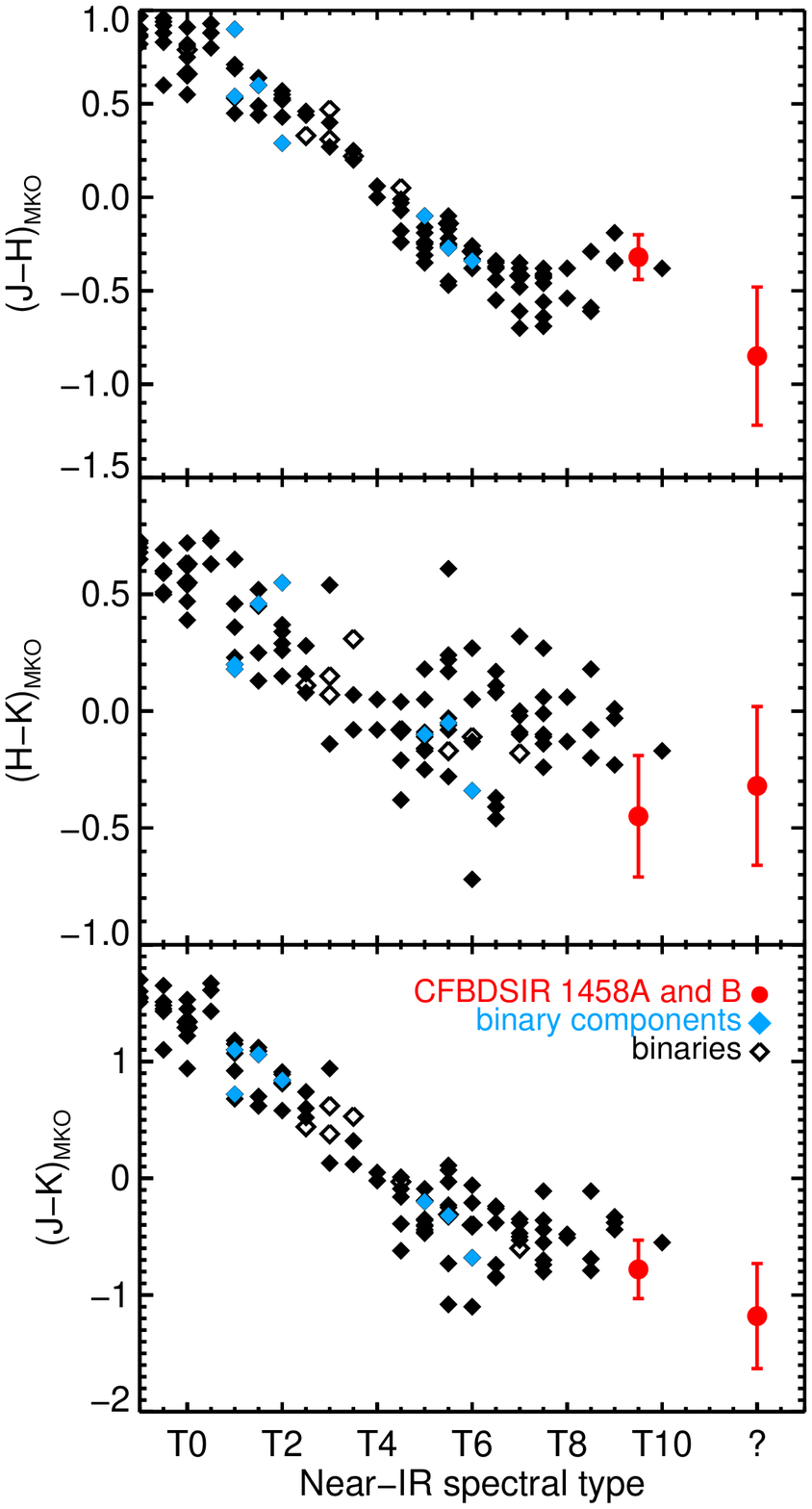}
\hskip -0.75in
\includegraphics[width=4in,angle=0]{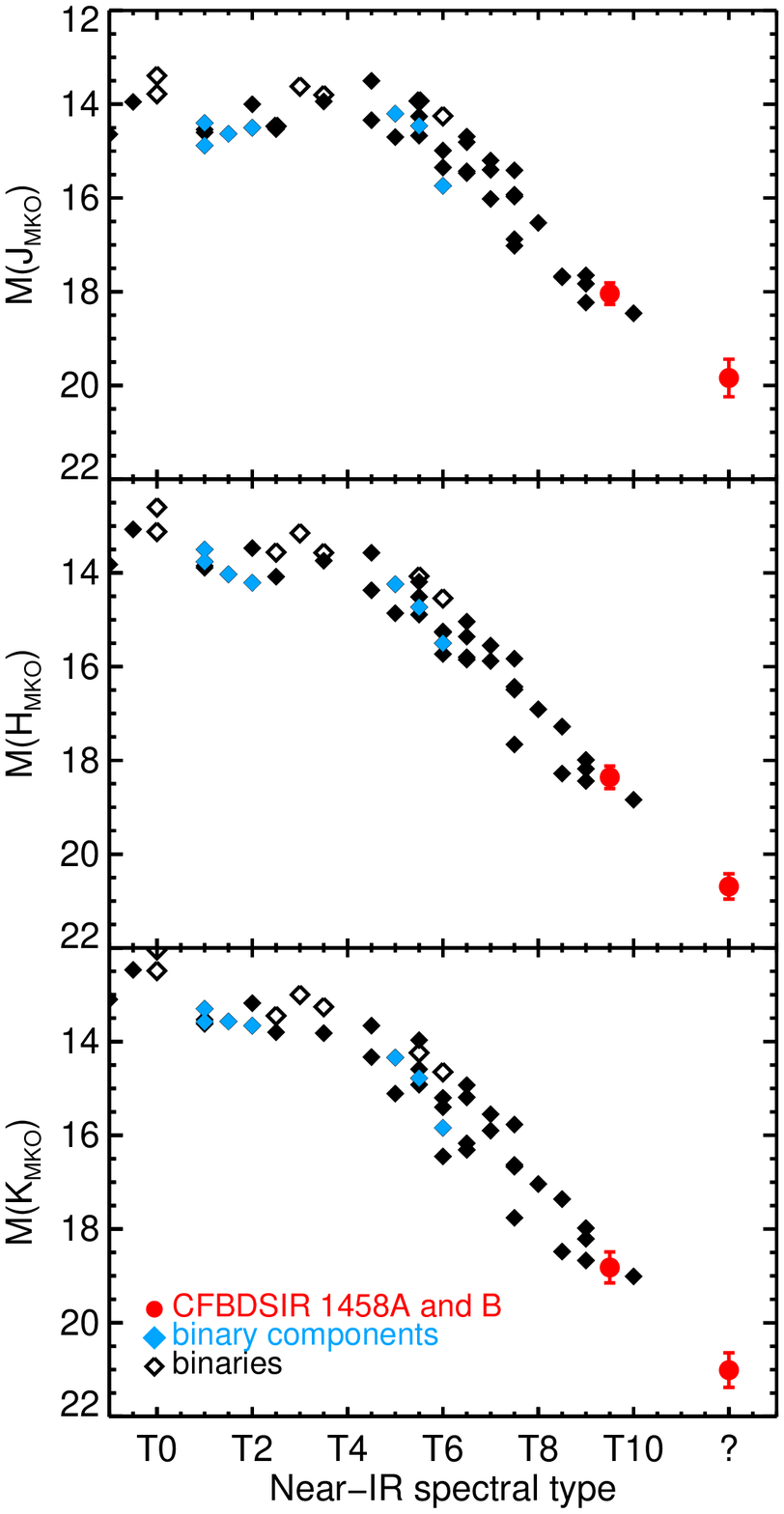}
%\includegraphics[width=4in,angle=0]{plot-properties-vs-spt-panel2.ps}
%\vskip 1ex
\caption{\normalsize Comparison of the colors and magnitudes of
  \cfbds{A} and {B} to those of field objects. The latter come from the
  compilation of \citet{2010ApJ...710.1627L}, augmented with resolved
  photometry of tight ultracool binaries
  \citep{2004A&A...413.1029M, 2005astro.ph..8082L, 2006astro.ph..5037L}.
  \label{fig:properties}}
\end{figure}

\begin{figure}
\centerline{\includegraphics[width=6in,angle=0]{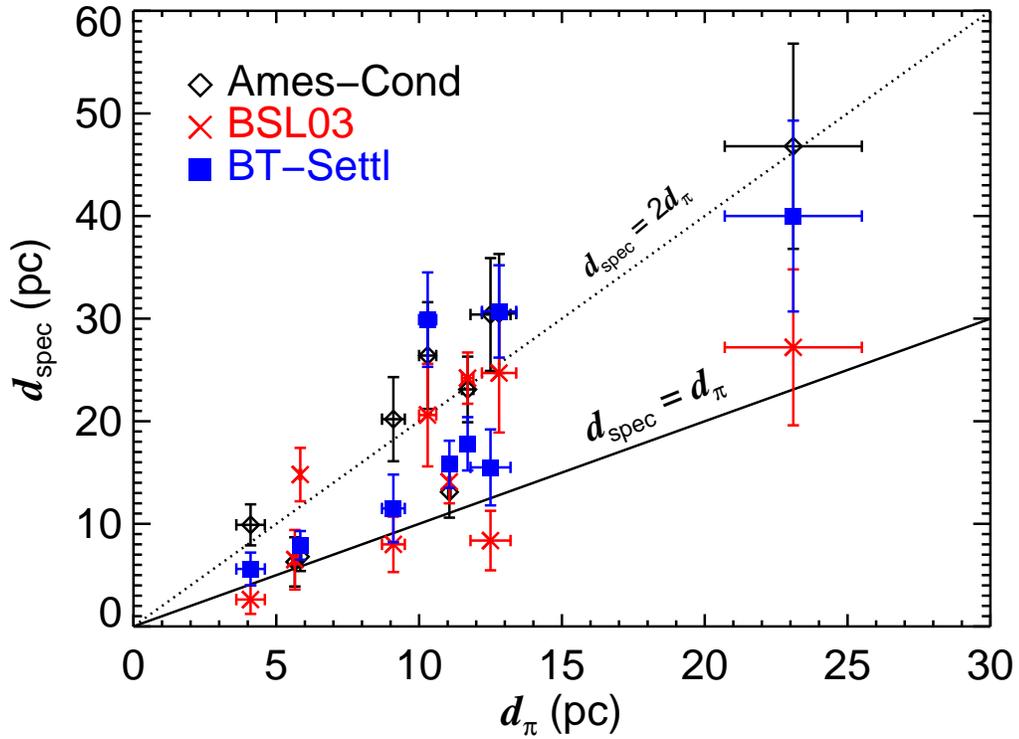}}
%\centerline{\includegraphics[width=6in,angle=0]{trig_par_distplot.eps}}
\vskip -4ex
\caption{\normalsize Comparison of model-derived spectroscopic distances ($d_{spec}$)
  using the method of \citet{2009ApJ...706.1114B} with the parallactic
  distances ($d_\pi$) for the sample of late-T dwarfs in
  Table~\ref{table:atmmodels}.  See Appendix for details.
  \label{fig:specdist}}
\end{figure}

%%======================================================================%%
%% Tables should be submitted one per page, so put a \clearpage before

%% Two options are available to the author for producing tables:  the
%% deluxetable environment provided by the AASTeX package or the LaTeX
%% table environment.  Use of deluxetable is preferred.
%%

%% Three table samples follow, two marked up in the deluxetable environment,
%% one marked up as a LaTeX table.

%% In this first example, note that the \tabletypesize{}
%% command has been used to reduce the font size of the table.
%% Note also that the \label command needs to be placed 

%----------------------------------------------------------------------%
\clearpage
\begin{deluxetable}{lccccccc}
\tablecaption{Keck LGS AO Observations \label{table:keck}}
\tabletypesize{\small}
\rotate
\tablewidth{0pt}
%\tablecolumns{2}
\tablehead{
  \colhead{Date} &
  \colhead{Filter\tablenotemark{a}} &
  \colhead{Airmass} &
  \colhead{FWHM} &
% \colhead{Strehl ratio} &
  \colhead{Separation} &
  \colhead{Position angle} &
  \colhead{$\Delta$mag} \\
  \colhead{(UT)} &
  \colhead{} &
  \colhead{} &
  \colhead{(mas)} &
  \colhead{(mas)} &
  \colhead{(deg)} &
  \colhead{}
}

\startdata

2010-May-22  & $\methshort$   & 1.14 & $62\pm6$  & $108\pm6$  & $308\pm2$      & $2.17\pm0.12$  \\   
2010-Jul-08  & $J$            & 1.33 & $94\pm10$ & $110\pm11$ & $308\pm4$      & $1.8\pm0.4$    \\   % 
             & $H$            & 1.04 & $61\pm4$  & $111\pm3$  & $309.3\pm1.4$  & $2.33\pm0.14$  \\   % 
             & $K$            & 1.11 & $60\pm3$  & $122\pm6$  & $314\pm2$      & $2.2\pm0.2$    \\   % 

\enddata

\tablecomments{All photometry on the MKO system. The tabulated
  uncertainties on the imaging performance (FWHM and Strehl ratio)
  are the RMS of the measurements from the individual images.}
% \tablenotetext{b}{The tabulated errors are computed by appropriately
%   combining in quadrature: (1) the instrumental measurements from
%   fitting the images of the binary and (2) the overall uncertainties
%   in the NIRC2 pixel scale and orientation.  The errors in parentheses
%   represent the instrumental errors alone.  See \S~2 for details.}

\end{deluxetable}

%----------------------------------------------------------------------%
\clearpage
\begin{deluxetable}{ccccccccccccc}
\tablecaption{CFHT/WIRCAM $J$-band Astrometry of \cfbds\ \label{table:astrometry}}
\tabletypesize{\small}
\rotate
\tablewidth{0pt}
%\tablecolumns{2}
\tablehead{
  \colhead{$N_{\rm ep}$} &
  \colhead{$N_{\rm ref}$} &
  \colhead{Epoch~1} &
  \colhead{$\Delta{t}$ (yr)} &
  \colhead{$\Delta$AM} &
  \colhead{$N_{\rm fr} (N_{\rm cal})$} &
  \colhead{$\mu$ (mas/yr)} &
  \colhead{P.A. (\degree)} &
  \colhead{$\pi_{\rm rel}$ (mas)} &
  \colhead{$\Delta{\pi}$ (mas)} &
  \colhead{$\pi_{\rm abs}$ (mas)} &
  \colhead{$\chi^2$ (dof)}
}

\startdata
7  &  72  &  2009.42  &  1.07  &  0.03  &  68 (8)  &  $432\pm6$  &  $154.2\pm0.7$  &  $ 42.4\pm4.5$ &  \phn$0.9\pm0.1$  &  $43.3\pm4.5$  &  11.5 (9)  \\

\enddata

\tablecomments{ 
  $N_{\rm ep}$~=~Number of distinct observing epochs (i.e., nights);
  $N_{\rm ref}$~=~Number of reference stars used;
  Epoch~1~=~Date of our first WIRCam observation;
  $\Delta{t}$~=~Time baseline for the observations;
  $\Delta$AM~=~Maximum difference in airmass over all epochs;
  $N_{\rm fr}$~=~Total number of frames obtained (10--11 per epoch);
  $N_{\rm cal}$~=~Subset of reference stars used in the absolute calibration;
  $\Delta\pi$~=~Correction applied to the best-fit parallax to determine
  the absolute parallax ($\pi_{\rm abs} = \pi_{\rm rel} + \Delta\pi$).
}

\end{deluxetable}

%------------------------------------------------------------------------------------------%
% BNARY-CALC.IDL:  mliu@notebook-3.local Tue Dec 21 00:53:25 2010
\clearpage
\begin{deluxetable}{lcc}
\tablecaption{Properties of \cfbdsAB \label{table:binary}}
\tablewidth{0pt}
\tabletypesize{\scriptsize}
\tablehead{
\colhead{Property} &
\colhead{Component {A}} &
\colhead{Component {B}}
}

\startdata

Integrated-light spectral type                                &      \multicolumn{2}{c}{T9.5}                  \\
$d$ (pc)                                                      &      \multicolumn{2}{c}{23.1 $\pm$ 2.4}        \\
Projected separation (AU)                                     &      \multicolumn{2}{c}{2.6 $\pm$ 0.3}         \\
$v_{tan}$ (km/s)                                               &      \multicolumn{2}{c}{47 $\pm$ 5}            \\
 % - - - - - - - - - - - - - - - - %                        &                         &                        \\
$J$ (mag)                                                     &  19.86 $\pm$ 0.07       &  21.66 $\pm$ 0.34     \\
$CH_4s$ (mag)                                                 &  19.73 $\pm$ 0.10       &  21.90 $\pm$ 0.15     \\
$H$ (mag)                                                     &  20.18 $\pm$ 0.10       &  22.51 $\pm$ 0.16     \\
$K$ (mag)                                                     &  20.63 $\pm$ 0.24       &  22.83 $\pm$ 0.30     \\
   % - - - - - - - - - - - - - - - - %                        &                         &                       \\
$J-H$ (mag)                                                   &   $-$0.32 $\pm$ 0.12    &   $-$0.85 $\pm$ 0.37  \\
$J-K$ (mag)                                                   &   $-$0.78 $\pm$ 0.25    &   $-$1.18 $\pm$ 0.45  \\
$CH_4s-H$ (mag)                                               &   $-$0.45 $\pm$ 0.14    &   $-$0.61 $\pm$ 0.22  \\
$H-K$ (mag)                                                   &   $-$0.45 $\pm$ 0.26    &   $-$0.32 $\pm$ 0.34  \\
   % - - - - - - - - - - - - - - - - %                        &                         &                       \\
$M(J)$ (mag)                                                  &  18.04 $\pm$ 0.23       &  19.84 $\pm$ 0.40     \\
$M(CH_4s)$ (mag)                                              &  17.91 $\pm$ 0.24       &  20.08 $\pm$ 0.27     \\
$M(H)$ (mag)                                                  &  18.36 $\pm$ 0.24       &  20.69 $\pm$ 0.27     \\
$M(K)$ (mag)                                                  &  18.82 $\pm$ 0.33       &  21.02 $\pm$ 0.37     \\
   % - - - - - - - - - - - - - - - - %                        &                         &                       \\ 
Near-IR spectral type                                         &       T9.5              &      $>$T10           \\
   % - - - - - - - - - - - - - - - - %                        &                         &                       \\
$\log(\Lbol/\Lsun)$ (dex)                                     &    $-6.02\pm0.14$       &    $-6.74\pm0.19$     \\

%----------------------------------------------------------------------%
% Lyon/COND: from Lbol, using CALC-MASSES.IDL, 2 ages
%----------------------------------------------------------------------%
\cutinhead{Evolutionary model results for 1.0 and 5.0~Gyr: Lyon/COND models and \Lbol} 
Mass ($M_{Jup}$)                                                     &   12.1 $\pm$ 1.9, 31 $\pm$ 4        &   5.8 $\pm$ 1.3,  14 $\pm$ 3 \\
Total mass ($M_{Jup}$)                                               &  \multicolumn{2}{c}{18 $\pm$ 2, 45 $\pm$ 5}         \\
$q\ (\equiv M_B/M_A)$                                               &   \multicolumn{2}{c}{0.48 $\pm$ 0.14, 0.45 $\pm$ 0.12}         \\
$T_{eff} (K)$                                                        &   556 $\pm$  48,  605 $\pm$  55      &    360 $\pm$  40,  380 $\pm$  50 \\
$\log(g)$ (cgs)                                                      &  4.45 $\pm$ 0.07,  5.00 $\pm$ 0.08  & 4.10 $\pm$ 0.10,  4.58 $\pm$ 0.11  \\
Orbital period (yr)                                                  &   \multicolumn{2}{c}{35$^{+28}_{-10}$, 22$^{+18}_{-6}$}          \\

%----------------------------------------------------------------------%
% Burrows97: from Lbol, using CALC-MASSES.IDL, 2 ages
%----------------------------------------------------------------------%
\cutinhead{Evolutionary model results for 1.0 and 5.0~Gyr: Burrows \etal\ (1997) models and \Lbol}  
Mass ($M_{Jup}$)                                                     &  13 $\pm$ 2,  36 $\pm$ 4         &    6.8 $\pm$ 1.5,  17 $\pm$ 4   \\
Total mass ($M_{Jup}$)                                              &  \multicolumn{2}{c}{20 $\pm$ 3, 53 $\pm$ 6}              \\
$q\ (\equiv M_B/M_A)$                                               &   \multicolumn{2}{c}{0.53 $\pm$ 0.15, 0.47 $\pm$ 0.13}    \\
$T_{eff} (K)$                                                        &  550 $\pm$  50,  600 $\pm$ 60    &     350 $\pm$ 40,  380 $\pm$ 50  \\
$\log(g)$ (cgs)                                                      &  4.47 $\pm$ 0.07, 5.06 $\pm$ 0.07    &   4.14 $\pm$ 0.10,  4.65 $\pm$ 0.12   \\
Orbital period (yr)                                                  &   \multicolumn{2}{c}{33$^{+27}_{-9}$, 20$^{+17}_{-6}$}       \\

%----------------------------------------------------------------------%
% BSL03: from M(J), using CALC-MASSES.IDL, 2 ages
%----------------------------------------------------------------------%
\cutinhead{Evolutionary model results for 1.0 and 5.0~Gyr: Burrows \etal\ (2003) models and $M(J)$}
Mass ($M_{Jup}$)                                                     &  11.1 $\pm$ 0.7, $>$25     &   7.6 $\pm$ 0.6,   18.8 $\pm$ 1.3  \\
Total mass ($M_{Jup}$)                                               &  \multicolumn{2}{c}{18.7 $\pm$ 0.9, $>$44}              \\
$q\ (\equiv M_B/M_A)$                                               &   \multicolumn{2}{c}{0.68 $\pm$ 0.07, $<$0.75}    \\
$T_{eff} (K)$                                                        &   479 $\pm$  20, $>$483      &    386 $\pm$  15,  407 $\pm$  15  \\  
$\log(g)$ (cgs)                                                      &   4.37 $\pm$ 0.03,  $>$4.85  &    4.19 $\pm$ 0.04,  4.69 $\pm$ 0.03  \\
Orbital period (yr)                                                  &   \multicolumn{2}{c}{34$^{+28}_{-10}$, $<$22}       \\
% %----------------------------------------------------------------------%

\enddata

\tablecomments{All infrared photometry on the MKO photometric system.
  For the 5~Gyr \citet{2003ApJ...596..587B} evolutionary model fits,
  component~A is brighter in $M(J)$ than the model grid, so its derived
  physical parameters are listed as limits.}

\end{deluxetable}

%----------------------------------------------------------------------------------------------------%
\clearpage
\begin{deluxetable}{ l c c c c c c c c c}
\tablecaption{Spectral indices for very late-T dwarfs.\label{table:indices}}     
\rotate
\tablewidth{0pt}
\tabletypesize{\tiny}
\tablehead{
  \colhead{Object}   &
  \colhead{Sp. Type} &
  \colhead{H$_2$O-J} &
  \colhead{W$_J$}    &
  \colhead{CH$_4$-J} &
  \colhead{H$_2$O-H} &     
  \colhead{CH$_4$-H} &
  \colhead{NH$_3$-H} &
  \colhead{CH$_4$-K} &
  \colhead{K/J}
}

\startdata                %        H2O-J               W_J                    CH4_J               H2O-H                CH4-H                NH3-H               CH4-K                K/J

Wolf 940B        & T8.5    & 0.030 $\pm$ 0.005  & 0.274 $\pm$ 0.002   & 0.1521 $\pm$ 0.0009& 0.142 $\pm$ 0.002  & 0.090 $\pm$ 0.002  & 0.537 $\pm$ 0.002 & 0.072 $\pm$ 0.013 & 0.1113 $\pm$ 0.0014\\  
ULAS J0034$-$00  & T8.5    & 0.012 $\pm$ 0.006  & 0.262 $\pm$ 0.004   & 0.145 $\pm$ 0.009  & 0.133 $\pm$ 0.010  & 0.097 $\pm$ 0.006  & 0.510 $\pm$ 0.011 & 0.092 $\pm$ 0.015 & 0.128 $\pm$ 0.003  \\  
CFBDS J0059$-$01 & T8.5    & 0.029 $\pm$ 0.005  & 0.257 $\pm$ 0.004   & 0.165 $\pm$ 0.005  & 0.119 $\pm$ 0.008  & 0.084 $\pm$ 0.002  & 0.526 $\pm$ 0.005 & 0.128 $\pm$ 0.037 & 0.101 $\pm$ 0.002  \\ 
\cfbds{AB}       & T9.5    &$-$0.002 $\pm$ 0.008& 0.237 $\pm$ 0.006   & 0.103 $\pm$ 0.009  & 0.078 $\pm$ 0.016  & 0.044 $\pm$ 0.018  & 0.54 $\pm$ 0.02   & 0.06 $\pm$ 0.19   & 0.058 $\pm$ 0.012  \\ 
\ugps\           & T10     & 0.0339 $\pm$ 0.0017& 0.2074 $\pm$ 0.0012 & 0.1358 $\pm$ 0.0019& 0.1218 $\pm$ 0.0017&0.0643 $\pm$ 0.0013 & 0.492 $\pm$ 0.002 & 0.096 $\pm$ 0.002 & 0.1262 $\pm$ 0.0003\\ 
\enddata

\tablecomments{Near-IR spectral indices and provisional spectral types
  for some of the latest-type brown dwarfs \citep{2007MNRAS.381.1400W,
    burningham08-T8.5-benchmark, 2008A&A...482..961D,
    2010MNRAS.408L..56L}. The index definitions are from
  \citet{2006ApJ...637.1067B}, \citet{2007MNRAS.381.1400W}, and
  \citet{2008A&A...482..961D}. Note that the negative value of H$_2$O-J
  for \cfbds\ is due to the fact that there is no detected flux in the
  1.14--1.165~\micron\ water absorption band probed by this index.}

\end{deluxetable}

%----------------------------------------------------------------------------------------------------%
\clearpage
\thispagestyle{empty}

\begin{deluxetable}{lccccccccccccccccc}
\rotate
%\tabletypesize{\tiny}
\tabletypesize{\scriptsize}
\setlength{\tabcolsep}{0.05in} 
\tablewidth{0pt}
\tablecolumns{17}
\tablecaption{Atmospheric Model Fits to Late-T Dwarfs with Parallaxes\label{table:atmmodels}}
\tablehead{
\colhead{}  & \colhead{}  & \multicolumn{4}{c}{Ames-Cond} & \colhead{} & \multicolumn{4}{c}{BSL03} &  \colhead{} &  \multicolumn{4}{c}{BT-Settl-2009} &  \colhead{} &  \colhead{} \\

\cline{3-6}
\cline{8-11}   
\cline{13-16}   

\colhead{}   &  \colhead{}   &  
\colhead{$T_\mathrm{eff}$}  & \colhead{log $g$}  &  \colhead{$R$\tablenotemark{a}} & \colhead{$d_\mathrm{spec}$ ($\sigma_\mathrm{mc}$, $\sigma_\mathrm{tot}$)\tablenotemark{b}} & 
\colhead{} & 
\colhead{$T_\mathrm{eff}$}  & \colhead{log $g$}  &  \colhead{$R$\tablenotemark{c}}  & \colhead{$d_\mathrm{spec}$ ($\sigma_{\mathrm mc}$, $\sigma_\mathrm{tot}$)\tablenotemark{b}}   & 
\colhead{} & 
\colhead{$T_\mathrm{eff}$}  & \colhead{log $g$}  &  \colhead{$R$\tablenotemark{c}}  & \colhead{$d_\mathrm{spec}$ ($\sigma_{\mathrm mc}$, $\sigma_\mathrm{tot}$)\tablenotemark{b}}   & 
\colhead{$d_{\pi}$ ($\sigma_{\pi}$)} &  \colhead{}  \\  

\colhead{Name}  & \colhead{Ref}  &  \colhead{(K)}  & \colhead{(cgs)}  &  \colhead{(R$_\mathrm{Jup}$)}  & \colhead{(pc)}     &
\colhead{} & 
\colhead{(K)}  & \colhead{(cgs)}  &  \colhead{(R$_\mathrm{Jup}$)}  & \colhead{(pc)}    & 
\colhead{} & 
\colhead{(K)}  & \colhead{(cgs)}  &  \colhead{(R$_\mathrm{Jup}$)}  & \colhead{(pc)}    & 
\colhead{(pc)} &  \colhead{$\pi$ Ref}   }

\startdata

Gl 570D (T7.5)        & 1    &  800 & 4.5  &  1.06  & 6.8 (0.1, 1.4)   &   & 797\tablenotemark{d}  &  4.80\tablenotemark{d}  & 0.996   &  14.8 (0.2, 2.6)  &   &  900  &  5.0  &  0.903  & 7.9  (0.2, 1.4) &  5.84 (0.03)\tablenotemark{e} &  13 \\
HD 3651B (T7.5)       & 2, 3 &  800 & 4.5  &  1.06  & 13.1 (0.2, 2.5)  &   & 797\tablenotemark{d}  &  4.80\tablenotemark{d}  & 0.996   &  14.1 (0.2, 2.1)  &   &  850  &  4.5  &  1.07   & 15.8 (0.2, 2.3) & 11.06 (0.04) & 13 \\
2MASS~J0415$-$09 (T8) & 4    &  800 & 5.0  &  0.89  & 6.29 (0.09, 2.4) &   & 703                   &  4.07                   & 1.33    &  6.50 (0.1, 2.9)  &   &  600  &  5.5  &  \nodata\tablenotemark{f} & \nodata\tablenotemark{f} & 5.65 (0.09) & 14 \\
Ross 458C (T8.5)      & 5, 6 &  900 & 4.0  &  1.24  & 23.1 (0.09, 3.2) &   & 859\tablenotemark{d}  &  4.24\tablenotemark{d}  & 1.27    &  24.2 (0.1, 2.5)  &   &  850  &  4.5  &  1.07   & 17.8 (0.1, 2.6) & 11.7 (0.2) & 13 \\
Wolf~940B (T8.5)      & 7    &  800 & 4.5  &  1.06  & 30.4 (0.3, 5.5)  &   & 483\tablenotemark{d}  &  4.85\tablenotemark{d}  & 0.924   &  8.37 (0.07, 2.9) &   &  650  &  5.0  &  0.874  & 15.5 (0.1, 3.7) & 12.5 (0.7)  & 15 \\
CFBDS J0059$-$01 (T9) & 8    &  700 & 4.5  &  1.04  & 20.2 (0.3, 4.1)  &   & 483\tablenotemark{d}  &  4.85\tablenotemark{d}  & 0.924   &  8.01 (0.1, 2.7)  &   &  600  &  5.0  &  0.879  & 11.5 (0.2, 3.3) &  9.2 (0.4)  & 16 \\
ULAS J0034$-$00 (T9)  & 9    &  800 & 4.5  &  1.06  & 30.5 (0.5, 5.8)  &   & 686  &  4.66  & 1.04  &    24.7 (0.4, 5.8)  &     &   750\tablenotemark{d}  &  4.0\tablenotemark{d} &  1.17  & 30.7 (0.5, 4.5)  & 12.8 (0.6) & 16, 17 \\
ULAS J1335+11 (T9)    & 10   &  800 & 4.5  &  1.06  & 26.4 (0.2, 5.2)  &   & 686  &  4.66  & 1.04  &    20.6 (0.1, 5.0)  &     &   800\tablenotemark{d}  &  4.0\tablenotemark{d} &  1.21  & 29.9 (0.1, 4.6)  & 10.3 (0.3) & 16 \\
\cfbds{AB} (T9.5)\tablenotemark{g} & 11 & 700 & 4.5 & 1.04 & 46.8 (1.0, 10) & & 555\tablenotemark{d} & 4.83\tablenotemark{d} & 0.946 & 27.2 (0.9, 7.6)  & &   700  &  5.0  &  0.880  &  40.0  (0.8, 9.3) & 23.1 (2.4) & 18 \\     % using resolved Jmag
\ugps\ (T10)          & 12   &  700 & 4.5  &  1.04  & 9.9 (0.1, 2.0)  &   & 421\tablenotemark{d}  &  4.72\tablenotemark{d}  & 0.961 &  2.62 (0.02, 1.4) &    &    600   &  5.0 &  0.879  & 5.6 (0.1, 1.6)   & 4.1 (0.5)  & 12 \\

\enddata

\tablecomments{Best-fitting Ames-Cond \citep{2001ApJ...556..357A}, BSL03
  \citep{2003ApJ...596..587B}, and BT-Settl-2009
  \citep{2010HiA....15..756A} model spectra to the combined
  1.15--1.35~$\mu$m, 1.45--1.6~$\mu$m, and 1.95--2.25~$\mu$m regions of
  the observed spectra.}

\tablenotetext{a}{Radius derived from an interpolated grid of Lyon/Cond
  evolutionary models. This value corresponds to the best-fitting
  $T_\mathrm{eff}$ and log $g$ Ames-Cond atmospheric model.}

\tablenotetext{b}{Model-derived spectroscopic distance estimate based on
  the method of \citet{2009ApJ...706.1114B}. $\sigma_{\mathrm mc}$
  denotes the Monte Carlo-derived distance error associated with the
  best-fitting atmospheric model and incorporates measurement and
  flux-calibration errors. $\sigma_\mathrm{tot}$ is the Monte
  Carlo-derived total measurement error in the distance, based on the
  measurement errors and assuming 1$\sigma$ uncertainties of 50~K and
  0.25~dex in $T_\mathrm{eff}$ and log $g$ in the model atmosphere
  fits.}

\tablenotetext{c}{Radius computed using the best-fitting
  $T_\mathrm{eff}$ and log $g$ and the power law fit to
  $R(T_\mathrm{eff}$, log $g)$ from Equation~1 of
  \citet{2003ApJ...596..587B}.}

\tablenotetext{d}{Best-fit model is at the edge of the model grid.}
\tablenotetext{e}{Distance to the primary Gl 570 A.}
\tablenotetext{f}{The best-fitting model atmosphere parameters are
  outside the grid of evolutionary models, so there is no corresponding
  radius or distance estimate.}
\tablenotetext{g}{Integrated light spectrum.}

\tablerefs{1--\citet{2000ApJ...531L..57B};
  2--\citet{2006astro.ph..8484M}; 3--\citet{2006astro.ph..9464L};
  4--\citet{burg01};
  5--\citet{2010MNRAS.405.1140G}; 6--\citet{2010A&A...515A..92S};
  7--\citet{burningham08-T8.5-benchmark}; % Wolf 940B: discovery
  8--\citet{2008A&A...482..961D}; 
  9--\citet{2007MNRAS.381.1400W};
  10--\citet{2008MNRAS.391..320B}; 
  11--\citet{2010A&A...518A..39D};
  12--\citet{2010MNRAS.408L..56L}; 
  13--\citet{2007A&A...474..653V};
  14--\citet{2004AJ....127.2948V}  % 2m 0415: parallax  
  15--\citet{1980AJ.....85..454H}  % Wolf 940B: parallax Harrington & Dahn
  16--\citet{2010A&A...524A..38M}; 
  17--\citet{2010A&A...511A..30S}; 
  18--this work.
}

\end{deluxetable}

%------------------------------------------------------------------------------------------------------------------------------------------------------%
\clearpage
\begin{deluxetable}{lccccccccccc}
\tabletypesize{\scriptsize}
\tablewidth{0pt}
\tablecolumns{12}
\tablecaption{Physical Properties Derived from Model Atmospheres and Parallactic Distances\label{table:tempmass}}
\tablehead{
\colhead{}  & \multicolumn{3}{c}{Ames-Cond} & \colhead{} & \multicolumn{3}{c}{BSL03} & \colhead{} & \multicolumn{3}{c}{BT-Settl-2009}  \\
\cline{2-4}
\cline{6-8}  
\cline{10-12}
\colhead{}     &  \colhead{$T_\mathrm{eff}$} & \colhead{log $g$} & \colhead{$M$} & \colhead{} & \colhead{$T_\mathrm{eff}$}& \colhead{log $g$}   & \colhead{$M$}  & \colhead{} & \colhead{$T_\mathrm{eff}$}  & \colhead{log $g$} & \colhead{$M$}  \\  
\colhead{Name} &  \colhead{(K)}             & \colhead{(cgs)}   & \colhead{(M$_\mathrm{Jup}$)} & \colhead{} & \colhead{(K)} & \colhead{(cgs)} & \colhead{(M$_\mathrm{Jup}$)}   & \colhead{} & \colhead{(K)}  & \colhead{(cgs)}  & \colhead{(M$_\mathrm{Jup}$)}  }

\startdata

Gl 570D (T7.5)          &  780--880 &  4.7--5.4 & 21--54 & & $>$520 & $>$4.4 & $>$13  & &  760--880  &  4.7--5.4 & 21--54 \\
HD 3651B (T7.5)         &  780--880 &  4.7--5.4 & 21--54 & & $>$700 & $>$4.6 & $>$22  & &  740--860  &  4.7--5.4 & 20--54 \\
2MASS J0415$-$09 (T8)   &  790--800 &  4.7--5.3 & 23--53 & & $>$660 & $>$4.6 & $>$19  & &  600--720  &  4.5--5.2 &  9--45 \\
Ross 458C (T8.5)        &  690--760 &  4.0--4.7 &  6--20 & & $>$590 & $>$3.9 & $>$5   & &  660--740  &  4.0--4.7 &  6--20 \\ 
%Ross 458C (T8.5)       &  750--840 &  4.7--5.4 & 21--55 & & $>$680 & $>$4.6 & $>$20  & &  720--840  &  4.7--5.4 & 20--55 \\ 
Wolf 940B (T8.5)        &  570--680 &  4.5--5.2 & 12--43 & & $>$500 & $>$4.4 & $>$12  & &  560--640  &  4.4--5.2 & 12--42 \\
CFBDS J0059$-$01 (T9)   &  520--600 &  4.4--5.2 & 10--35 & & $>$440 & $>$4.3 & $>$10  & &  500--600  &  4.4--5.1 & 11--34 \\
ULAS J0034$-$00 (T9)    &  560--680 &  4.5--5.2 & 12--43 & & $>$500 & $>$4.4 & $>$12  & &  560--660  &  4.4--5.2 & 12--42 \\
ULAS J1335+11 (T9)      &  560--660 &  4.4--5.2 & 12--42 & & $>$500 & $>$4.4 & $>$12  & &  560--640  &  4.4--5.2 & 12--42 \\
\cfbds{AB} (T9.5)\tablenotemark{a} & 520--640 & 4.4--5.2 & 11--42 & & $>$460 & $>$4.3  &  $>$10 & & 520--620 & 4.4--5.2 & 11--36 \\    % using resolved Jmag
%\cfbds{AB}\tablenotemark{a} & 540--660 & 4.4--5.2 & 12--42 & & $>$480 & $>$4.3  &  $>$10 & & 540--660 & 4.4--5.2 & 12--42 \\   % using int-light Jmag
\ugps\ (T10)            &  480--600 &  4.4--5.1 & 10--35 & & $>$420 & $>$4.3 &  $>$9  & &  480--580  &  4.4--5.1 & 10--33 \\

\enddata
\tablenotetext{a}{Integrated light spectrum.}
\end{deluxetable}

\end{document}